\documentclass[12pt,american]{article}
\usepackage[T1]{fontenc}
\usepackage{geometry}
\usepackage{color}
\usepackage{babel}
\usepackage{enumitem}
\usepackage{amsmath}
\usepackage{amsthm}
\usepackage{amssymb}
\usepackage{stmaryrd}
\usepackage{setspace}
\usepackage[authoryear]{natbib}
\usepackage[unicode=true,pdfusetitle,
 bookmarks=true,bookmarksnumbered=false,bookmarksopen=false,
 breaklinks=false,pdfborder={0 0 1},backref=false,colorlinks=false]
 {hyperref}
\hypersetup{
 colorlinks,linkcolor={blue!70!black},citecolor={blue!50!black},urlcolor={blue!60!black}}

\makeatletter
\theoremstyle{remark}
\newtheorem{rem}{\protect\remarkname}
\theoremstyle{definition}
\newtheorem{defn}{\protect\definitionname}
\theoremstyle{plain}
\newtheorem{thm}{\protect\theoremname}
\theoremstyle{plain}
\newtheorem{lem}{\protect\lemmaname}
\theoremstyle{plain}
\newtheorem{cor}{\protect\corollaryname}

\usepackage{lmodern}
\usepackage{setspace}
\usepackage{babel}
\usepackage{float}
\usepackage{mathrsfs}
\usepackage{breakurl}
\usepackage{bbm}
\usepackage{tikz}

\allowdisplaybreaks

\makeatother

\providecommand{\corollaryname}{Corollary}
\providecommand{\definitionname}{Definition}
\providecommand{\lemmaname}{Lemma}
\providecommand{\remarkname}{Remark}
\providecommand{\theoremname}{Theorem}

\begin{document}
\global\long\def\a{\alpha}%
 
\global\long\def\b{\beta}%
 
\global\long\def\g{\gamma}%
 
\global\long\def\d{\delta}%
 
\global\long\def\e{\epsilon}%
 
\global\long\def\l{\lambda}%
 
\global\long\def\t{\theta}%
 
\global\long\def\o{\omega}%
 
\global\long\def\s{\sigma}%

\global\long\def\G{\Gamma}%
 
\global\long\def\D{\Delta}%
 
\global\long\def\L{\Lambda}%
 
\global\long\def\T{\Theta}%
 
\global\long\def\O{\Omega}%
 
\global\long\def\R{\mathbb{R}}%
 
\global\long\def\N{\mathbb{N}}%
 
\global\long\def\Q{\mathbb{Q}}%
 
\global\long\def\I{\mathbb{I}}%
 
\global\long\def\P{\mathbb{P}}%
 
\global\long\def\E{\mathbb{E}}%
\global\long\def\B{\mathbb{\mathbb{B}}}%
\global\long\def\S{\mathbb{\mathbb{S}}}%
\global\long\def\V{\mathbb{\mathbb{V}}\text{ar}}%

\global\long\def\X{{\bf X}}%
\global\long\def\cX{\mathscr{X}}%
 
\global\long\def\cY{\mathscr{Y}}%
 
\global\long\def\cA{\mathscr{A}}%
 
\global\long\def\cB{\mathscr{B}}%
\global\long\def\cF{\mathscr{F}}%
 
\global\long\def\cM{\mathscr{M}}%
\global\long\def\cN{\mathcal{N}}%
\global\long\def\cG{\mathcal{G}}%
\global\long\def\cC{\mathcal{C}}%
\global\long\def\sp{\,}%

\global\long\def\es{\emptyset}%
 
\global\long\def\mc#1{\mathscr{#1}}%
 
\global\long\def\ind{\mathbf{\mathbbm1}}%
\global\long\def\indep{\perp}%

\global\long\def\any{\forall}%
 
\global\long\def\ex{\exists}%
 
\global\long\def\p{\partial}%
 
\global\long\def\cd{\cdot}%
 
\global\long\def\Dif{\nabla}%
 
\global\long\def\imp{\Rightarrow}%
 
\global\long\def\iff{\Leftrightarrow}%

\global\long\def\up{\uparrow}%
 
\global\long\def\down{\downarrow}%
 
\global\long\def\arrow{\rightarrow}%
 
\global\long\def\rlarrow{\leftrightarrow}%
 
\global\long\def\lrarrow{\leftrightarrow}%

\global\long\def\abs#1{\left|#1\right|}%
 
\global\long\def\norm#1{\left\Vert #1\right\Vert }%
 
\global\long\def\rest#1{\left.#1\right|}%

\global\long\def\bracket#1#2{\left\langle #1\middle\vert#2\right\rangle }%
 
\global\long\def\sandvich#1#2#3{\left\langle #1\middle\vert#2\middle\vert#3\right\rangle }%
 
\global\long\def\turd#1{\frac{#1}{3}}%
 
\global\long\def\ellipsis{\textellipsis}%
 
\global\long\def\sand#1{\left\lceil #1\right\vert }%
 
\global\long\def\wich#1{\left\vert #1\right\rfloor }%

\global\long\def\abs#1{\left|#1\right|}%
 
\global\long\def\norm#1{\left\Vert #1\right\Vert }%
 
\global\long\def\rest#1{\left.#1\right|}%
 
\global\long\def\inprod#1{\left\langle #1\right\rangle }%
 
\global\long\def\ol#1{\overline{#1}}%
 
\global\long\def\ul#1{\underline{#1}}%
 
\global\long\def\td#1{\tilde{#1}}%

\global\long\def\upto{\nearrow}%
 
\global\long\def\downto{\searrow}%
 
\global\long\def\pto{\overset{p}{\longrightarrow}}%
 
\global\long\def\dto{\overset{d}{\longrightarrow}}%
 
\global\long\def\asto{\overset{a.s.}{\longrightarrow}}%

\title{A Partial Order on Preference Profiles\thanks{I thank Yi Chen, Shaowei Ke, George Mailath, Xiaosheng Mu, Andrew
Postlewaite, Larry Samuelson, Yiqing Xing and Chen Zhao for helpful
discussions, comments and suggestions. I also thank Simon Oros for
capable research assistance.}}
\author{Wayne Yuan Gao\thanks{Gao: Department of Economics, University of Pennsylvania. 133 South
36th Street, Philadelphia, PA 19104, USA, waynegao@upenn.edu. }}
\maketitle
\begin{abstract}
\noindent We propose a theoretical framework under which preference
profiles can be meaningfully compared. Specifically, we first define
the ranking vector of an allocation as the vector of all individuals\textquoteright{}
rankings of this allocation. We then induce a partial order on preference
profiles from a partial order on the ranking vectors of Pareto efficient
allocations. We characterize the maximal and minimal elements under
this partial order. In particular, we illustrate how an individualistic
form of social preferences can be maximal in a specific setting. We
also discuss how the framework can be further generalized to incorporate
additional economic ingredients.
\end{abstract}
\newpage{}

\section{\label{sec:Intro}Introduction}

Preferences, and preference profiles as the collection of the preferences
held by a set of individuals, are often times the primitives of an
economic model, and they also usually form the basis for the evaluation
of economic outcomes. 

However, in real life, preferences themselves are often discussed
and compared; furthermore, sometimes people may even hold strong opinions
on such comparisons. For an example of considerable social importance,
consider the advance of LGBT rights, along with the accompanying public
debates and evolving social attitudes towards the LGBT community \citep{belmonte2020international}. 

In addition, even though preferences are typically taken to be exogenously
given in most economic analysis, at least to some degree preferences
can be shaped within a society through a variety of social environments,
such as families, schools, companies and governments: see \citet*{postlewaite2011social}
a survey on the social determinants of preferences. A natural question
to ask then is: if an individual, an entity or the society as a whole
can choose what kind of preferences (and consequently preference profiles)
to induce in a target population, then in what formal sense can one
say that a certain type of preference (profile) is ``better than''
an alternative type on a normative ground?

While \emph{by no means} this paper can capture the sophistication
of these complex but important social topics, we seek to propose a
theoretical framework for a potentially meaningful comparison of preference
profiles, though admittedly on a very stylized and abstract level.

~

Specifically, we propose a partial order on preference profiles. Given
a fixed (finite) set of feasible allocations $X$, and any preference
profile $P$ on $X$, we first define the Pareto frontier $PE_{X}\left(P\right)$
under $P$ and $X$. For each Pareto efficient allocation $x\in PE_{X}\left(P\right)$,
we then define each individual $i$'s ranking $R_{X}\left(P_{i},x\right)$
of $x$ among all allocations in $X$ under $P_{i}$. We then define
a partial order on preference profiles based on the ranking evaluations
of Pareto frontiers. We say one preference profile $P$ is ``\emph{more
Pareto-favorable than}'' another  preference $P^{'}$, and write
``$P\trianglerighteqslant_{X}P^{'}$'', if there exists an \textit{onto}
mapping $\psi$ from the Pareto frontier of $P^{'}$ onto the Pareto
frontier of $P$, such that the ranking vector of any Pareto efficient
allocation under $P^{'}$ is weakly dominated by the ranking vector
of the image allocation under $\psi$ :
\[
R_{X}\left(P,\psi\left(x\right)\right)\leq R_{X}\left(P^{'},x\right),\quad\forall x\in PE_{X}\left(P^{'}\right).
\]
We now provide a discussion about this proposed partial order, explain
why it may be sensible, and discuss its difference from related literature.

First, the proposed partial order on preference profiles is defined
purely based on \textit{intra-personal} comparisons of a given individual's
own ranking evaluations according to two candidate preferences of
her own. It should be emphasized that there is no whatsoever \textit{inter-personal}
comparison or aggregation of preferences, and hence this paper is
\textit{not} an analysis of social welfare, social justice or social
equality, which have been extensively studied in welfare economics,
dating back to the heated debates in the pioneering work by \citet*{rawls1958justice},
\citet*{harsanyi1975can} and \citet*{binmore1989}. Conceptually,
welfare economics is primarily concerned with the selection of some
allocations, or in certain scenarios a single allocation, from the
set of all feasible allocations, under a \emph{single} given preference
profile. This paper adopts Pareto efficiency as such a solution concept,
but is primarily concerned with comparisons across \emph{different
}preference profiles. 

Second, the proposed partial order on preference profiles does not
reflect the ``\textit{preference}'' of any external entities, but
only encodes the preferences of the set of involved individuals themselves.
Under a ``weakly more Pareto-favorable'' preference profile, everyone
may obtain a weakly higher-ranked allocation according to their own
preferences on the Pareto frontier, relative to what they may obtain
in the Pareto frontier of a ``weakly less Pareto-favorable'' preference
profile. The weak ranking improvement holds for every individual in
the society.

Third, the comparison across different preference profiles via the
proposed partial order is anchored by two invariant objects: the fixed
set of feasible allocations $X$\textit{, }and the \emph{ordinal}
structure of preferences on $X$. For a given individual, two distinct
preferences of this individual are specified on exactly the same set
of objects, and given preferences are ordinal, all preferences are
completely characterized by the relative ranking of different allocations
in $X$. Suppose that we \emph{have to} make a systematic comparison
across different preference profiles, then we have to rely on \emph{some}
\emph{i}nvariant structures of preference profiles as a lever of comparison,
and in this case the use of ranking seems at least natural, if not
necessary under some axiomatic foundations.

Forth, conceptually the analysis in this paper can be thought as a
study of monotone comparative statics in an inverted direction: we
induce a partial order on preference profiles based on a partial order
on the set of Pareto efficient allocations, and then seek to characterize
the extremal elements (as well as intermediate transitions) under
the induced partial order. The techniques underlying the analysis
in this paper are combinatorial in nature, and is in particular based
on the standard approach for monotone comparative statics that exploit
lattice theory and supermodularity \citep{topkis1998supermodularity,milgrom1994monotone}. 

Lastly, as should be clear from the discussion above, this paper does
not move outside the paradigm of \textit{economic rationality}, but
only explores a potential framework that compares preference profiles,
taking as given the economic outcomes that may arise from each given
preference profile. In this paper, we use Pareto efficiency as a solution
concept for economic outcomes, but other solution concepts based on
efficiency or equilibrium considerations may be of interest in different
scenarios. The individuals in this paper have full rationality with
respect to their common knowledge of the preference profile, the associated
Pareto frontier and their optimal choices. Hence, this paper itself
does not fall into, or seek to address, the debate between rationalistic
economics and behavioral economics or neuroeconomics, represented
by \citet*{gul2008case}, \citet*{camerer2008case} and \citet*{hausman2008mindless}
in \citet*{caplin2008foundations}. 

~

The rest of the paper is organized as following. Section \ref{sec:Ex_2A2A}
illustrates the key idea underlying the proposed partial order via
a simple two-individual two-allocation example. Section \ref{sec:GenAbstract}
lays out the model setup in a general finite allocation space with
general preferences, and characterize the maximal and minimal elements
under the proposed partial order in the space of all possible preference
profiles. Section \ref{subsec:Str_Private} considers a setting where
both the allocation space and the preference profiles are endowed
with some sensible structures: specifically, we consider the assignment
of one and only one widget among a discrete set of widgets to each
individual, where each individual's preference is private in the sense
that she only cares about her own widget assignment. Section \ref{subsec:Str_Private}
contains an illustrative example with continuous allocation space,
and we conclude in Section \ref{sec:Conclusion} with a discussion
about how the framework can be further generalized to incorporate
additional economically relevant modeling ingredients. The proofs
are available in the Appendix.

\section{\label{sec:Ex_2A2A}A Two-Individual Two-Allocation Example}

We start with a simple example with two individuals and two feasible
allocations to illustrate the key idea of this paper.

Specifically, let ``1'' and ``2'' denote two individuals, and
``$r$'', ``$b$'' denote two pens: ``$r$'' for a red pen and
``$b$'' for a blue pen. Consider the problem of allocating the
two pens to the two individuals, under the assumption that each of
the two individuals must be allocated one and only pen. Then, the
set of admissible allocations, denoted by $X$, may be written denoted
as
\[
X:=\left\{ \left(r,b\right),\left(b,r\right)\right\} ,
\]
where ``$\left(r,b\right)$'' stands for ``allocate the red pen
to individual 1, and the blue pen to individual 2'', and ``$\left(b,r\right)$''
the vice versa.

Consider two possible strict ordinal preference profiles $P=\left(P_{1},P_{2}\right)$
and $P^{'}=\left(P_{1}^{'},P_{2}^{'}\right)$ on the set of admissible
allocations $X$, where ``$P_{i}$'' and $\text{\textquotedblleft}P_{i}^{'}\text{\textquotedblright}$
stand for individual $i$'s preferences. Suppose that under the first
preference profile $P$, individual $1$ prefers the red pen over
the blue pen, while individual $2$ prefers the blue pen over the
red pen (and neither individual intrinsically cares about what pen
the other individual is getting), inducing a preference profile on
$X$ defined by:
\begin{equation}
P:\quad\left(r,b\right)\succ_{P_{1}}\left(b,r\right),\quad\left(r,b\right)\succ_{P_{2}}\left(b,r\right),\label{eq:ex_P}
\end{equation}
where ``$\succ_{P_{i}}$'' denotes strict preference according to
$P_{i}$. Under the second preference profile $P^{'}$, suppose that
both individual 1 and individual 2 prefer getting the red pen over
the blue pen:
\begin{align}
P^{'} & :\quad\left(r,b\right)\succ_{P_{1}^{'}}\left(b,r\right),\quad\left(b,r\right)\succ_{P_{2}^{'}}\left(r,b\right).\label{eq:ex_Pprime}
\end{align}
The question we ask is: is there any \emph{precise} and \emph{sensible}
way to compare the two preference profiles $P$ and $P^{'}$? 

~

For this simple example, it is straightforward to see that, under
the first preference profile $P$, there is a unique Pareto efficient
allocation $\left(r,b\right)$, which allocates to each individual
her favorite pen. However, under the second preference profile $P^{'}$,
both allocations $\left(r,b\right)$ and $\left(b,r\right)$ in $X$
are Pareto efficient, and moreover there is at least one individual
who does not get her favorite pen. It is then at least \emph{intuitive}
to say that the first preference profile $P$ is ``better'' in the
sense that individuals are able to obtain their best possible allocation,
according to each of their own preferences in $P$, via \emph{market
exchanges}, while the same cannot be said for the second preference
profile $P^{'}$.

To represent the idea above mathematically, we first compute the sets
of Pareto efficient allocations in $X$ under $P$ and $P^{'}$, denoted
by $PE_{X}\left(P\right)$ and $PE_{X}\left(P^{'}\right)$:
\begin{align*}
PE_{X}\left(P\right)=\left\{ \left(r,b\right)\right\} ,\quad & PE_{X}\left(P^{'}\right)=\left\{ \left(r,b\right),\left(b,r\right)\right\} .
\end{align*}
Then, for each preference profile, we compute each individual's \emph{ranking}
of all Pareto efficient allocations in $X$ according to this given
individual's preference. Specifically, under $P$, we have only one
Pareto efficient allocation $\left(r,b\right)$, to which individual
$1$ assigns a ranking of $R_{X}\left(P_{1},\left(r,b\right)\right)=1$
according to individual $1$'s preference $P_{1}$ on $X$, and to
which individual 2 also assigns a ranking of $R_{X}\left(P_{2},\left(r,b\right)\right)=1$,
or in vector form:
\[
R_{X}\left(P,\left(r,b\right)\right)=\left(1,1\right).
\]
Similarly, under $P^{'}$, we compute the ranking vectors of the two
Pareto efficient allocations in $PE_{X}\left(P^{'}\right)$ as
\begin{align*}
R_{X}\left(P^{'},\left(r,b\right)\right)=\left(1,2\right),\quad & R_{X}\left(P^{'},\left(b,r\right)\right)=\left(2,1\right).
\end{align*}
In this example, the intuitive urge, if any, to consider $P$ as ``\emph{better
than}'' $P^{'}$, should be coming from the observation that all
Pareto efficient allocations under $P$ are assigned \emph{better
ranking vectors} than all Pareto efficient allocations under $P^{'}$:
\begin{equation}
R_{X}\left(P,x\right)\lneqq R_{X}\left(P^{'},y\right),\ \forall x\in PE_{X}\left(P\right),\ \forall y\in PE_{X}\left(P^{'}\right).\label{eq:ex_PEdom}
\end{equation}
We may define ``$P\vartriangleright_{X}P^{'}$'' based on \eqref{eq:ex_PEdom}.

~

We offer some immediate remarks. The first is on the comparison of
rankings across different preference profiles.
\begin{rem}[About the Use of Ranking]
\label{rem:ranking}Clearly, the definition of the comparison ``$P\vartriangleright_{X}P^{'}$''
relies crucially on the comparison of the ranking vectors across allocations
in $P$ and $P^{'}$ based on \eqref{eq:ex_PEdom}, and whether such
a comparison is meaningful or not at all is admittedly not a trivial
question. 

However, we argue that, for this simple two-individual two-allocation
example, the proposed comparison ``$P\vartriangleright_{X}P^{'}$''
has some nice features. First, we have not pre-imposed any form of
preference relations on the allocation set $X$ from outside the two
given preference profiles $P$ and $P^{'}$. Second, we have maintained
completely symmetric treatment of individuals beyond their potentially
heterogeneous preferences on $X$. Third, given the previous two points,
if we have to make a comparison between preference profiles, it seems
very hard to imagine a scenario where a preference profile like $P$,
under which there is a unique reasonable economic outcome allocation
(the single Pareto efficient allocation) that every individual finds
to be her own unique favorite allocation among the set of all admissible
allocations $X$, is not defined to the ``best possible'' preference
profile.

Admittedly, these are not precisely defined axioms that necessitate
the use of rankings as the ``invariant lever of comparison'' across
different preference profiles. However, it is hoped that this remark
provides some intuitive motivations for the use of rankings, and it
seems interesting for future research to explore the axiomatic approaches
that either support the use of rankings or suggest new definitions.
\end{rem}
\begin{rem}[About Pareto Dominance]
\label{rem:Pareto}It should be pointed out that the comparison between
two preference profiles $P$ and $P^{'}$ described above is not a
comparison about \emph{social welfare }per se. Recall that, in the
standard economics literature, a welfare comparison is made between
two admissible (feasible) allocations in the set of all admissible
(feasible) allocations $X$, under a \emph{single} given preference
profile $P$. The comparison ``$P\vartriangleright_{X}P^{'}$'',
however, involves two preference profiles $P$ and $P^{'}$. On the
other hand, the comparison ``$P\vartriangleright_{X}P^{'}$'' obviously
has tight relationship with \emph{Pareto dominance,} an important
concept in welfare economics. 
\end{rem}
First, the comparison ``$P\vartriangleright_{X}P^{'}$'' is based
on the Pareto efficient allocations only under each preference profile.
The focus on the Pareto efficient allocations is intended to impute,
or at least to represent, economic gains from any trade or exchange
opportunities under each given ordinal preference profile. On a more
abstract level, the set of Pareto efficient allocations, or the mapping
$PE_{X}$, represent one form of well-defined \emph{economic solution
concepts} defined for any given preference profile. In particular,
the solution concept of Pareto efficiency is based on a specific form
of efficiency criterion. More generally, one could substitute Pareto
efficiency $PE_{X}$ with any economic solution concept ${\cal E}_{X}$
that picks a set of \emph{economic outcome }allocations within $X$.
It is popular to select ${\cal E}_{X}$ either based on \emph{efficiency}
considerations or \emph{equilibrium} considerations in economics.
In this paper, we adopt Pareto efficiency as a representative solution
concept, not only because it is arguably the most robust form of efficiency
criteria, but also because that, relatedly, in many economic scenarios
the corresponding equilibrium outcomes are also Pareto efficient.

Second, the comparison ``$P\vartriangleright_{X}P^{'}$'' based
on \eqref{eq:ex_PEdom} is based on uniform (elementwise) inequality
between the ranking vectors taken from the two sets of Pareto efficient
allocations under two different preference profiles, which shares
some formal similarity with the definition of Pareto dominance. This
is intended to make the comparison ``$P\vartriangleright_{X}P^{'}$''
as robust as possible. In particular, notice that in the description
leading up to \eqref{eq:ex_PEdom}, we have not made any comparison
(or aggregation) of preference relations \emph{across} different individuals:
every individual's ranking of each admissible allocation is either
compared with her own ranking of another  admissible allocation under
the same preference profile, or compared with her own ranking of some
allocation under a different preference profile.

\section{\label{sec:GenAbstract}General Preferences on General Allocation
Space}

\subsection{\label{subsec:Gen_Setup}Setup and Definition}

Let $N:=\left\{ 1,...,N\right\} $ be a set of $N$ individuals (with
slight abuse of notation), and let $X:=\left\{ 1,...,M\right\} $
be a set of $M$ admissible or feasible allocations (or actions).
Let $P_{i}$ denote each individual $i$'s preference on $X$, with
``$x\succ_{P_{i}}x^{'}$'' denoting ``$i$ strictly prefers $x$
over $x^{'}$ according to $P_{i}$'', and ``$x\succsim_{P_{i}}x^{'}$''
and ``$x\sim_{P_{i}}x^{'}$'' denoting weak preference of $x$ over
$x^{'}$ and indifference between $x$ and $x^{'}$ according to $P_{i}$,
respectively. Let $P:=\left(P_{i}\right)_{i\in N}$ for the preference
profile. We say a preference profile $P$ is \emph{strict }if either
$x\succ_{P_{i}}x^{'}$ or $x^{'}\succ_{P_{i}}x$ holds for any $i\in N$
and any two distinct $x,x'\in X$. 

Given the feasibility set $X$ and a given individual $i$'s preference
$P_{i}$ on $X$, denote individual $i$'s ranking of a given allocation
$x\in X$ among all allocations in $X$ according to $i$'s own preference
$P_{i}$ as
\[
R_{X}\left(P_{i},x\right):=1+\#\left\{ z\in X:z\succ_{P_{i}}x\right\} ,\ \forall x\in X,
\]
i.e., $x$ is the $R_{X}\left(P_{i},x\right)$-th best allocation
in $X$ according to $i$'s preference $P_{i}$. We may write the
ranking vector of a given allocation $x$ under $P$ as
\[
R_{X}\left(P,x\right):=\left(R_{X}\left(P_{1},x\right),R_{X}\left(P_{2},x\right),...,R_{X}\left(P_{N},x\right)\right).
\]

Given the feasibility set $X$ and the preference profile $P$ on
$X$, denote the set of Pareto efficient allocations in $X$ under
the preference profile $P$ by
\begin{align*}
PE_{X}\left(P\right) & :=\left\{ x\in X:\ x\text{ is not strictly Pareto dominated by any }z\in X\right\} .
\end{align*}

~

We now propose the following definition of a partial order on preference
profiles, which generalizes the one described in Section \ref{sec:Ex_2A2A}
via \eqref{eq:ex_PEdom}. An important and necessary departure from
the two-individual two-allocation case is to accommodate the fact
that, in general, the ranking vector of an arbitrary Pareto efficient
allocation under $P$ may not be comparable to an arbitrary Pareto
efficient allocation under $P^{'}$. Our proposed adaption is to define
the partial order by the existence, or lack thereof, of a mapping
from $PE\left(P^{'}\right)$ to $PE\left(P\right)$ such that the
ranking vectors are comparable along the mapping in a consistent way.
\begin{defn}[A Partial Order $\trianglerighteqslant_{X}$ on Preference Profiles]
\label{def:def_order} Given a feasibility set $X$ and any two preference
profiles $P$ and $P^{'}$ defined on $X$, we say that $P$ is weakly
``\emph{more Pareto-favorable than}'' $P^{'}$ on $X$, and write
\[
P\trianglerighteqslant_{X}P^{'},
\]
whenever there exists an \emph{onto} mapping $\psi:PE_{X}\left(P^{'}\right)\to PE_{X}\left(P\right)$
such that
\begin{align*}
R_{X}\left(P,\psi\left(x\right)\right) & \leq R_{X}\left(P^{'},x\right),\quad\text{for any }x\in PE_{X}\left(P^{'}\right).
\end{align*}
We write ``$P\vartriangleright_{X}P^{'}$'' for the strict case,
where $P\trianglerighteqslant_{X}P^{'}$ but $P^{'}\ntrianglerighteqslant_{X}P$.
We write ``$P\triangleq_{X}P^{'}$'' for the case where $P\trianglerighteqslant_{X}P^{'}$
and $P^{'}\trianglerighteqslant_{X}P$.
\end{defn}
Clearly, $\trianglerighteqslant_{X}$ is a partial order. The requirement
of $\psi$ being \emph{onto }is to ensure that there is no allocation
$y\in PE_{X}\left(P\right)$ that does not dominate some $x\in PE_{X}\left(P^{'}\right)$
in terms of the ranking vector. It is easy to check that, for the
two-individual two-allocation example in Section \ref{sec:Ex_2A2A},
we may set the mapping $\psi$ by $\psi\left(x\right)=\left(r,b\right)$
for either $x\in PE_{X}\left(P^{'}\right)=\left\{ \left(r,b\right),\left(b,r\right)\right\} $.

~

For notational simplicity, in the following we will suppress the subscript
``$\cd_{X}$'' in ``$R_{X}$'', ``$PE_{X}$'' and ``$\trianglerighteqslant_{X}$'',
given that $X$ is a primitive of the analysis in this paper. We will
write out the subscript ``$\cd_{X}$'' explicitly when we emphasize
the dependence of the partial order on $X$.

~

There are potentially many other ways to define the partial order
as a generalization of the one proposed in the two-individual two-allocation
example. We discuss in the following remarks, as well as the corresponding
appendices, two alternative versions of definitions that are different
from but closely related to Definition \ref{def:def_order}.
\begin{rem}
\label{rem:def_v2}We may define an alternative version $\tilde{\trianglerighteqslant}$
of the partial order in the following way. We define a quasi order
$P\tilde{\vartriangleright}P^{'}$, if there exists an \emph{onto}
mapping $\psi:PE\left(P^{'}\right)\to PE\left(P\right)$ such that
\begin{align*}
R\left(P,\psi\left(x\right)\right) & \leq R\left(P^{'},x\right),\quad\text{for any }x\in PE\left(P^{'}\right),
\end{align*}
with at least one inequality being strict in the following sense:
\[
R\left(P_{i},\psi\left(x\right)\right)<R\left(P_{i}^{'},x\right),\quad\text{for some }i\in N\text{ and some }x\in PE\left(P^{'}\right).
\]
Note that $\tilde{\vartriangleright}$ and $\vartriangleright$, though
very similar, are not equivalent in general. In particular, note that
$\tilde{\vartriangleright}$ is stronger than $\vartriangleright$
in the sense that $P\tilde{\vartriangleright}P^{'}$ implies $P\vartriangleright P^{'}$
but not vice versa. See Appendix for more discussion and adapted results
on $\tilde{\vartriangleright}$.
\end{rem}

\subsection{\label{subsec:Gen_Extreme}Extremal Elements}

We now provide a characterization of the maximal and minimal elements
under the partial order $\trianglerighteqslant$ on the space of all
possible preference profiles on $X$.

We say that a preference profile $P$ is a $\trianglerighteqslant$\emph{-maximal
element} if $P^{'}\trianglerighteqslant P$ implies that $P^{'}\triangleq P$,
that $P$ is a $\trianglerighteqslant$\emph{-minimal element} if
$P\trianglerighteqslant P^{'}$ implies that $P\triangleq P^{'}$,
that $P$ is a $\trianglerighteqslant$\emph{-upper bound }if $P\trianglerighteqslant P^{'}$
holds for any $P^{'}$, and that $P$ is a $\trianglerighteqslant$\emph{-lower
bound }if $P^{'}\trianglerighteqslant P$ holds for any $P^{'}$. 
\begin{thm}[Extremal Elements under $\trianglerighteqslant$]
\label{thm:discrete_gen} Consider the space of all possible preference
profiles on $X$.
\begin{itemize}
\item[(a)]  Maximal elements and least upper bounds: $\ol P$ is a $\trianglerighteqslant$-maximal
element and a $\trianglerighteqslant$-upper bound if and only if
there exists a unique $x^{*}\in X$ such that all individuals like
$x^{*}$ the best under $\ol P$, i.e.,
\begin{equation}
R\left(\ol P\right)\left[x^{*}\right]={\bf 1}_{N}.\label{eq:allrank1}
\end{equation}
\item[(b)]  Minimal elements: $\ul P$ is a $\trianglerighteqslant$-minimal
element if and only if both of the following conditions hold:
\begin{itemize}
\item[(i)]  every allocation in $X$ is Pareto efficient under $\ul P$, i.e.,
\begin{equation}
PE\left(\ul P\right)=X.\label{eq:allPE}
\end{equation}
\item[(ii)] $\ul P$ is strict.
\end{itemize}
\end{itemize}
\end{thm}
Lower bounds do not exist in general. Example:
\[
P=\left(\begin{array}{ccc}
1 & 2 & 3\\
3 & 1 & 2\\
2 & 3 & 1
\end{array}\right),\quad P^{'}\left(\begin{array}{ccc}
1 & 2 & 3\\
3 & 1 & 2\\
3 & 2 & 1
\end{array}\right)
\]
$P$ and $P^{'}$ are both minimal elements as $PE\left(P\right)=PE\left(P^{'}\right)=X$.
However, no preference profile $P^{''}$ can be weakly dominated by
both, $P\trianglerighteqslant P^{''}$ and $P^{'}\trianglerighteqslant P^{''}$,
because at least one dominance must be strict, which is impossible
given the minimality of $P$ and $P^{'}$.

~

Theorem \ref{thm:discrete_gen}(a) is a direct generalization of the
intuition in Remark \ref{rem:ranking} for the two-individual two-allocation
example that, if everyone gets her unique favorite allocation in the
unique Pareto efficient allocation under a given preference profile,
then this preference profile should be considered as the ``best''.
The proof is almost trivial: there exists a trivial onto mapping $\psi$
from $PE\left(P\right)$ to $PE\left(\ol P\right)=\left\{ x^{*}\right\} $
that satisfies Definition \ref{def:def_order}, given that $R\left(\ol P,x^{*}\right)={\bf 1}_{n}$
achieves the best possible ranking vector. Note that Theorem \ref{thm:discrete_gen}(a)
does not say that $\ol P$ must be unique: in general there exist
many $\trianglerighteqslant_{X}$-maximal preference profiles that
are equivalent with each other under $\trianglerighteqslant_{X}$,
and all such preference profiles are also $\trianglerighteqslant_{X}$-upper
bounds.

Theorem \ref{thm:discrete_gen}(b), which characterizes the $\trianglerighteqslant_{X}$-minimal
elements, is also a direct generalization of the preference profile
$P^{'}$ defined in the two-individual two-allocation example, where
$PE\left(P^{'}\right)=\left\{ \left(r,b\right),\left(b,r\right)\right\} =X$.
Intuitively, more strictly conflicting preference relations on $X$
produce a larger set of Pareto efficient allocations, which would
make it harder for a given allocation to Pareto dominate another ,
which requires congruent preference between two allocations across
all individuals with at least one strict preference. Moreover, as
more preference relations become strict, ranking vectors tend to worsen:
with indifference, it might be possible to have two different allocations
$x,x^{'}$ that some individual $i$ finds to both be $i$'s favorite
allocation under some $P_{i}$, so that $R\left(P_{i},x\right)=R\left(P_{i},x^{'}\right)=1$;
however, breaking the indifference between $x$ and $x^{'}$ in $P_{i}$,
while keeping $i$'s all other preferences on $X$ unchanged, would
increase the ranking vector as it is no longer possible to have $R\left(P_{i},x\right)=R\left(P_{i},x^{'}\right)=1$.

Technically, the proof of Theorem \ref{thm:discrete_gen}(b) is involved,
especially the ``\emph{only if}'' part (with the proof of the ``\emph{if}''
part being relatively simple once we have established the ``\emph{only
if}'' part). Despite the technicality, the proof of the ``\emph{only
if}'' part is worth some more discussion, which is sketched through
three lemmas presented in the next subsection. These lemmas not only
serve to characterize the minimal elements under $\trianglerighteqslant_{X}$,
but also reveal how any non-minimal preference profile $P$ that does
not satisfy the condition in Theorem \ref{thm:discrete_gen}(b) can
be \emph{locally perturbed} to construct an alternative preference
$P^{'}$ such that $P\vartriangleright_{X}P^{'}$. This is informative
on comparisons between $P$ and $P^{'}$ even if neither $P$ or $P^{'}$
is maximal nor minimal.

\subsection{\label{subsec:Gen_ToMin}Intermediate Transitions}

In this subsection, we sketch the proof of the ``\emph{only if}''
part of Theorem \ref{thm:discrete_gen}(b) via three lemmas. Each
of the following lemmas is proved via the construction of an alternative
preference profile $P^{'}$ by ``locally perturbing'' a given preference
profile $P$ that is $\trianglerighteqslant_{X}$-minimal, followed
by a demonstration of $P\vartriangleright P^{'}$. In words, we show
how any non-minimal preference profile can be transformed towards
a $\trianglerighteqslant$-minimal preference profile step by step.

Lemma \ref{lem:AllPE} below, along with its proof, demonstrate how
a preference profile $P$ cannot be $\trianglerighteqslant$-minimal
if the set of Pareto efficient allocations under $P$ does not equal
the whole allocation space $X$.
\begin{lem}[Universal Pareto Efficiency for Minimality]
\label{lem:AllPE}If $PE\left(P\right)\neq X$, then $P$ is not
$\trianglerighteqslant$-minimal.
\end{lem}
The detailed proof is available in the Appendix, but here we briefly
describe how to construct an alternative preference profile $P^{'}$
such that $P\vartriangleright P^{'}$ when $PE\left(P\right)\neq X$.

Specifically, we may take any Pareto dominated allocation $\ul x\in X\backslash PE\left(P\right)$,
which is Pareto dominated by some $\ol x\in PE\left(P\right)$. Moreover,
we show that we can always select the allocation $\ol x$ to be such
that some individual $\ol i\in N$ strictly prefers $\ol x$ over
$\ul x$, and moreover $\ol x$ is individual $\ol i's$ favorite
allocation among all Pareto efficient allocations that Pareto dominate
$\ul x$.

We then construct another  preference profile $P^{'}$ by switching
individual $\ol i$'s rankings of $\ul x$ and $\ol x$ in $P$ only,
i.e., setting
\[
R\left(P_{\ol i}^{'},\ul x\right):=R\left(P_{\ol i},\ol x\right),\quad R\left(P_{\ol i}^{'},\ol x\right):=R\left(P_{\ol i},\ul x\right),
\]
while keeping individual $\ol i$'s rankings of all other allocations,
as well as all other individuals' rankings of all allocations, completely
unchanged from $P$. Then it can be shown that $P\vartriangleright P^{'}$.

Under the newly constructed preference profile $P^{'}$, the allocation
$\ul x$, which is not Pareto efficient under the original preference
profile $P$, becomes Pareto efficient: $\ul x\in PE\left(P^{'}\right)$.
Intuitively, notice that the ranking vector for $\ul x$ is weakly
improved uniformly from $P$ to $P^{'}$, so any allocation that does
not Pareto dominate $\ul x$ under $P$ cannot Pareto dominates $\ul x$
under $P^{'}$, either. Moreover, individual $\ol i$'s ranking vector
for $\ul x$ is strictly improved to such an extent that individual
$\ol i$ now ranks $\ul x$ better under $P^{'}$ than all other Pareto
efficient allocations that Pareto dominate $\ul x$ under $P$. Hence,
$\ul x$ becomes Pareto efficient under $P^{'}$. For this new Pareto
efficient allocation $\ul x$, we can set the function $\psi$ in
Definition \ref{def:def_order} to map $\ul x\in PE\left(P^{'}\right)$
to $\ol x\in PE\left(P\right)$, so that the ranking vector of $\ul x$
under $P^{'}$ must still be no better than the ranking vector of
$\psi\left(\ul x\right)=\ol x$ under $P$, because the rankings of
all individuals other than $\ol i$ have not changed. If $\ol x$
remains Pareto efficient, we could also set $\psi$ to map $\ol x$
to itself, noting that the ranking vector of $\ol x$ strictly worsens
from $P^{'}$ to $P$, due to individual $\ol i$'s switch of rankings
between $\ol x$ and a strictly inferior $\ul x$ from $P$.

The above summarizes the key intuition for why $P\vartriangleright P^{'}$,
but of course the set of Pareto efficient allocations may involve
other changes that we have not discussed above. Moreover, we also
need to make sure that the mapping $\psi$ can be fully configured
in an appropriate way. The formal proof in Appendix \ref{subsec:pf_lem_AllPE}
contains all these details.

~

Lemmas \ref{lem:No22indiff} and \ref{lem:AllStrict} below, along
with their proofs, demonstrate how a preference profile $P$ cannot
be $\trianglerighteqslant$-minimal if the preference $P$ is not
strict in two specific manners.
\begin{lem}[Partially Strict Preferences for Minimality]
\label{lem:No22indiff}If there exist any two distinct individuals
$i,j\in N$ and any Pareto efficient allocation $x\in PE\left(P\right)$
such that $x\sim_{P_{i}}y$ and $x\sim_{P_{j}}y$ for some $y\neq x$,
then $P$ is not $\trianglerighteqslant$-minimal.
\end{lem}
\begin{lem}[Strict Preferences for Minimality]
\label{lem:AllStrict} If there exists a Pareto efficient allocation
$x\in PE\left(P\right)$ such that $x\sim_{P_{i}}y$ for some individual
$i$ and some allocation $y\neq x$, then $P$ is not $\trianglerighteqslant$-minimal.
\end{lem}
Essentially, Lemma \ref{lem:AllStrict} states that, for $\trianglerighteqslant$-minimality,
there cannot be any indifferences in $P$ that involve any Pareto
efficient allocation. The restriction to comparisons involving Pareto
efficient allocations is very intuitive, given that the partial order
$\trianglerighteqslant$ is defined based on comparisons of (ranking
vectors of) Pareto efficient allocations.

Lemma \ref{lem:No22indiff} is less interpretable than Lemma \ref{lem:AllStrict},
but it is only stated as an intermediate step to prove Lemma \ref{lem:AllStrict}.
Even though the proof of Lemma \ref{lem:No22indiff} is rather tedious,
the underlying idea is quite simple: if we have two individuals $ij$
and two allocations $x,y$ such that $x\sim_{P_{i}}y$ and $x\sim_{P_{j}}y$,
we can perturb the preference profile $P$ by ``breaking the indifferences''
for individuals $i$ and $j$ in \emph{opposite directions}, i.e.,
setting $P^{'}$ to ensure $x\succ_{P_{i}^{'}}y$ and $y\succ_{P_{j}^{'}}x$.
In the meanwhile, we can keep all other preference relations unchanged,
so that this perturbation essentially only affects the ranking between
$x$ and $y$.\footnote{Technically, there can be other allocations $z$ such that $z\sim_{P_{i}}x$
or $z\sim_{P_{j}}x$. See the proof in Appendix \ref{subsec:pf_lem_no22}
for how such kind of allocations are handled.} This perturbation should have increased the ranking vectors uniformly
across all allocations and all individuals. Of course, there are many
more technical subtleties beyond this simple intuition: see the proof
in Appendix \ref{subsec:pf_lem_no22} for details.

The proof of Lemma \ref{lem:AllStrict} is much simpler once Lemma
\ref{lem:No22indiff} is proved: essentially, we can just ``break
the indifference'' between $x$ and $y$ for individual $i$. See
the proof in Appendix \ref{subsec:pf_lem_strict} for details.

\section{\label{sec:Structure}Preferences on Structured Allocation Space}

The previous section considers general preference profiles on an arbitrary
finite set of feasible allocation space. However, we may often be
interested in some forms of allocations that have meaningful structures,
under which certain preference profiles may be automatically excluded
from considerations. This section is thus intended to provide an illustration
on how the key idea underlying Definition \ref{def:def_order} can
be flexibly adapted to accommodate and exploit sensible primitive
structures or restrictions built in the space of admissible allocations
as well as the space of admissible preference profiles.

\subsection{\label{subsec:Str_Private}Private Preferences}

Consider the allocation of $M$ distinct indivisible widgets to a
group of $N$ individuals, where each individual is assigned one and
only one widget.\footnote{We may take one of the widgets as a ``null widget'', but we do not
consider this case for simplicity.)} Formally, we write the set of widgets as $M:=\left\{ 1,...,M\right\} ,$
and the set of admissible allocations as
\[
X:=\left\{ x\in M^{n}:\#\left\{ i:x_{i}=k\right\} \leq1,\ \forall k\in M\right\} .
\]

Suppose that each individual's preference is \emph{private} in the
sense that each agent only cares about the widget assigned to herself.
Formally, each individual $i$'s \textit{private} preference is characterized
by a \emph{strict} preference $P_{i}$ on the set of widgets $M$.
Given any allocation $x\in X$, individual $i$'s ranking of $x$
under $P_{i}$ is dependent on $x_{i}$ only:
\begin{align*}
R_{X}\left(P_{i},x\right) & \equiv R_{M}\left(P_{i},x_{i}\right):=1+\#\left\{ k\in M:k\succ_{P_{i}}x_{i}\right\} .
\end{align*}
Note that, even though $P_{i}$ is a strict preference on $M$, individual
$i$'s implied preference on $X$ is not strict: conditional on getting
a given widget $x_{i}\in M$, individual $i$ is indifferent across
all allocations $z\in X$ such that $z_{i}=x_{i}$.

The definition of the partial order $\trianglerighteqslant$ is still
given in Definition \ref{def:def_order}.

It should be clear that the specification described here is a more
structured generalization of the two-individual two-pen example in
Section \ref{sec:Ex_2A2A}. Notice in particular that each individual's
preference is restricted to be private, and thus constrained to allow
for indifferences in a structured way. Hence, Theorem \ref{thm:discrete_gen}
does not apply directly: in this section we would like to respect
the structures laid out above, and focus on making partial order comparisons
only among preference profiles that satisfy the structures imposed
this section.
\begin{thm}[Extremal Elements among Private Preferences in Structured Allocation
Space]
\label{thm:private} Suppose $N\leq M$.
\begin{itemize}
\item[(a)]  Maximal elements: A preference profile $\ol P$ is $\trianglerighteqslant_{X}$-maximal
if and only if $\ol P$ is such that there exists $N$ distinct widgets
$\ol x_{1},...,\ol x_{n}$ in $M$ such that $R\left(\ol P_{i},\ol x_{i}\right)=1$
for all $i\in N$, i.e. everyone's favorite widget in $M$ is different. 
\item[(b)]  Minimal elements: A preference profile $P$ is $\trianglerighteqslant_{X}$-minimal
if and only if $\ul P$ is such that there exists a subset $M_{N}$
of $M$ such that (i) all individuals rank the widgets in $M_{N}$
to be no worse than their $N$-th best widget in $M$, i.e., $R\left(\ul P_{i},z\right)\leq N$
for all $i\in N$ and all $z\in M_{N}$, and (ii) all individuals
preferences restricted on $M_{n}$ coincide, i.e., $\rest{\ul P_{i}}_{M_{N}}=\rest{\ul P_{j}}_{M_{N}}$
for all $i,j\in N$.
\item[(c)]  Extremal elements are also bounds: All preference profiles $\ol P$
are $\succ_{X}$-upper bounds, and all preference profiles $\ul P$
are $\succ_{X}$-lower bounds.
\end{itemize}
\end{thm}
Under the current setting, Theorem \ref{thm:private} provides a formal
foundation for the desirability of \emph{diversity, }or \emph{individual
heterogeneity}, in private preferences. Specifically, $\trianglerighteqslant$-maximality
is characterized by full diversity in the top choices across individual
private preferences, while $\trianglerighteqslant$-minimality is
characterized by full alignment of the top-$N$ choices in individual
private preferences. 

These results are highly consistent with the results in Theorem \ref{thm:discrete_gen},
despite the dissimilarities in appearance. For maximality, Theorem
\ref{thm:private}(a) and Theorem \ref{thm:discrete_gen}(a) are equivalent
after accounting for the additional structures in the current setting.
For minimality, though it is in general no longer possible to find
a preference profile $P$ such that $P$ is strict and $PE\left(P\right)=X$
in the current specification, the full conformity of private preferences
on the common top-$N$ widgets under $\ul P$ in Theorem \ref{thm:private}(b)
induces an effectively \emph{strict} preference profile over the set
of allocations consisting of the top-$N$ widgets only, which also
coincides with the set of Pareto efficient allocations under $\ul P$.

However, compared with Theorem \ref{thm:discrete_gen} in the last
section, Theorem \ref{thm:private} further exploits the imposed structure
built into the allocation space and the space of admissible preference
profiles, thus gaining more specificity. It is plausible that, in
many economic scenarios, the allocation space can be factorized as
a product space of individual-specific allocations subject to some
budget constraints over total allocations, and moreover in many scenarios
individuals are more concerned with their own widget allocation rather
than what widgets other individuals get. At least in such scenarios,
Theorem \ref{thm:private} is more relevant or interesting than Theorem
\ref{thm:discrete_gen}.

~

The proof of Theorem \ref{thm:private} utilizes the following sequence
of lemmas and corollaries, each of which is interpretable and intuitive.
\begin{lem}[No Cycles of Envy]
\label{lem:NoCycle} For any $x\in PE\left(P\right)$, if there are
a sequence of distinct individuals $i_{1},...,i_{k}$ such that $x_{i_{h+1}}\succ_{P_{i_{h}}}x_{i_{h}}$for
every $h=1,...,k-1,$ we must have $x_{i_{k}}\succ_{P_{ik}}x_{i_{1}}$.
\end{lem}
The proof is simple. Otherwise, an obvious trading cycle would constitute
a strict Pareto improvement. Note that the above holds in particular
for pairs, i.e., $x_{j}\succ_{P_{i}}x_{i}\imp x_{j}\succ_{P_{j}}x_{i}.$
\begin{lem}[No Better Widgets Available]
\label{lem:NotAvail} For any $x\in PE\left(P\right)$, any $i\in N$
and any $k$ such that $k<R\left(P_{i},x_{i}\right),$ there must
exists some other $j\in N$ with $R\left(P_{i},x_{j}\right)=k.$
\end{lem}
The proof is again simple. Suppose not. Then the individual can find
an available widget that she likes better, resulting in a Pareto improvement.

An intuitive but useful corollary follows immediately from Lemma \ref{lem:NotAvail}: 
\begin{cor}
For any $x\in PE\left(P\right)$ and any $i\in N$, $R\left(P_{i},x_{i}\right)\leq N.$
\end{cor}
The next lemma that characterizes a form of upper bounds on the ranking
vector of a Pareto efficient allocation.
\begin{lem}[Bounds on Ranking Vectors of Pareto Efficient Allocations]
\label{lem:Diagonal}For any $x\in PE\left(P\right)$ and any $k\in\left\{ 1,....,N\right\} $,
there are at least $k$ individuals who are allocated widgets weakly
better than their $k$-th favorite widgets:
\[
\#\left\{ i\in N:\ R\left(P_{i},x_{i}\right)\leq k\right\} \geq k.
\]
\end{lem}
\begin{cor}
\label{cor:DiagonalComp}There are at most $\left(N-k+1\right)$ individuals
who are allocated widgets weakly worse than their $k$-th favorite
widgets, i.e., $\#\left\{ i\in N:\ R\left(P_{i},x_{i}\right)\geq k\right\} \leq N-k+1.$
\end{cor}
Based on Lemma \ref{lem:Diagonal} and Corollary \ref{cor:DiagonalComp},
we can deduce that the ranking vector $R\left(P,x\right)$ of any
Pareto efficient allocation $x\in PE\left(P\right)$ can be bounded
above by a permutation of the vector $\left(1,2,...,N\right)$. It
can also be shown that this upper bounded is attainable, and the type
of preference profiles that attain this upper bound are $\trianglerighteqslant$-minimal.
See the proofs in Appendices \ref{subsec:pf_lem_diag} and \ref{subsec:pf_thm_private}
for details.

\subsection{\label{subsec:Str_Social}Social Preferences}

Section \ref{subsec:Str_Private} considers a setting where individuals
have strict private preferences over their own widgets but they are
completely indifferent over the widgets other individuals obtain.
In this section, we seek to incorporate social preferences, where
each individual not only cares about the widget she gets herself,
but may also care about which widgets other individuals get. However,
if we do not impose any structure on social preferences, we are effectively
back in a setting with general preferences over a general allocation
space, as already considered in Section \ref{sec:GenAbstract}. To
investigate partially the middle ground between Section \ref{subsec:Str_Private}
and \ref{sec:GenAbstract}, we introduce the following specification.

The set of admissible allocations $X$ is again characterized as the
assignment of widgets to individuals, as described in Section \ref{subsec:Str_Private}.
We assume that each individual $i$'s preference $P_{i}$ can be factorized
into two components $P_{i}=\left(P_{i,i},P_{i,-i}\right)$ where $P_{i,i}$
is a well-defined \emph{strict} preference relation on $M$ while
$P_{i,-i}$ is a well-defined preference relation on $X_{-i}:=\left\{ x_{-i}=\left(x_{j}\right)_{j\neq i}:x\in X\right\} .$
Moreover, we require that $P_{i}$ satisfy the following \emph{lexicographic}
structure:
\begin{itemize}
\item[(L1)] $x\succ_{P_{i}}y$ if $x_{i}\succ_{P_{i,i}}y_{i}$, or if $x_{i}=y_{i}$
and $x_{-i}\succ_{P_{i,-i}}y_{-i}$.
\item[(L2)]  $x\sim_{P_{i}}y$ if $x_{i}=y_{i}$ and $x_{-i}\sim_{P_{i,-i}}y_{-i}$.
\end{itemize}
The lexicographic structure of $P_{i}$ is an extreme modeling device
to induce the plausible feature that individuals are primarily concerned
with their own widget allocations. The factorization of individual
preference and the lexicographic structure jointly allow us to essentially
separate the analysis of the \emph{private} component $P_{i,i}$ of
each individual $i$'s preference on her own widget from the \emph{social
}component $P_{i,-i}$ of her preference on other individuals' widgets,
lending great analytical tractability while conveying the key intuition.
\begin{defn}
\label{def:social_pref} Given a preference profile $P$ on $X$ and
any individual $i\in N$, we say that $P_{i,-i}$ is ``\textit{\textcolor{black}{individualistic}}''
if, for any $x_{-i},y_{-i}\in X_{-i}$, we have $x_{-i}\sim_{P_{i,-i}}y_{-i}$.
\end{defn}
Clearly, an individual $i$ with an \textit{\textcolor{black}{individualistic}}
$P_{i,-i}$ is only concerned with her own widget allocation, but
is completely indifferent with respect to other individuals' widgets,
just as in Section \ref{subsec:Str_Private}. In particular, every
individual holds no prejudice against another individual's enjoyment
of private widgets, even if the definitions of ``good widgets''
differ dramatically across individuals. 

Given any strict private preference profile $\left(P_{i,i}\right)_{i\in N}$,
and any arbitrary social preference profile $\left(P_{i,-i}\right)_{i\in N}$,
we first observe that the set of Pareto efficient allocations under
the preference profile $P:=\left(P_{i,i},P_{i,-i}\right)_{i\in N}$
is invariant with respect to the social preference profile $\left(P_{i,-i}\right)_{i\in N}$
due to the imposed lexicographic structure. 

Formally, define 
\begin{equation}
P^{ind}:=\left(P_{i,i},P_{i,-i}^{ind}\right)_{i\in N}\label{eq:P_ind}
\end{equation}
and we have the following lemma.
\begin{lem}[Invariance of Pareto Efficient Allocations]
\label{lem:CarryOver} $PE\left(P\right)=PE\left(P^{ind}\right)$.
\end{lem}
Lemma \ref{lem:CarryOver} allows us to be free of concerns about
differences in the sets of Pareto efficient allocations, but focus
purely on the ranking vectors. Hence, we may carry over all results
from Section \ref{subsec:Str_Private} about the sets of Pareto efficient
allocations.

The following theorem shows that $P^{ind}$ is $\trianglerighteqslant$-maximal
among preference profiles that share the same profile of private preferences. 
\begin{thm}[$\trianglerighteqslant$-Maximality of Individualism]
\label{thm:IndMax}For any $P=\left(P_{i,i},P_{i,-i}\right)_{i\in N}$
and $P^{ind}=\left(P_{i,i},P_{i,-i}^{ind}\right)_{i\in N}$, we have:
\begin{itemize}
\item[(a)] $P^{ind}\trianglerighteqslant P$.
\item[(b)] $P^{ind}\vartriangleright P$ if there exist some $i\in N$ and two
distinct Pareto allocations $x,y\in PE\left(P\right)$ such that $x_{i}=y_{i}$
and $y_{-i}\succ_{P_{i,-i}}x_{-i}$.
\item[(c)] $P^{ind}\vartriangleright P$ if $N\geq3$, $\#\left(PE\left(P\right)\right)\geq2$
and $\left(P_{i,-i}\right)_{i\in N}$ is strict.
\end{itemize}
\end{thm}
The proof of Theorem \ref{thm:IndMax}(a) is simple. Consider any
Pareto efficient allocation $x\in PE\left(P\right)$. Each individual
$i$'s ranking of $x$ in $X$ is given by:

\begin{align*}
R\left(P_{i},x\right) & =1+\#\left\{ z\in X:z_{i}\succ_{P_{i}}x\right\} \\
 & =1+\#\left\{ z\in X:z_{i}\succ_{P_{i,i}}x_{i}\right\} +\#\left\{ z\in X:z_{i}=x_{i}\text{ and }z_{-i}\succ_{P_{i,-i}}x_{-i}\right\} \\
 & \geq1+\#\left\{ z\in X:z_{i}\succ_{P_{i,i}}x_{i}\right\} \\
 & =1+\#\left\{ z\in X:z_{i}\succ_{P_{i}^{ind}}x_{i}\right\} \\
 & =R\left(P_{i}^{ind},x\right),
\end{align*}
where the equality holds if and only if
\begin{equation}
\#\left\{ z\in X:z_{i}=x_{i}\text{ and }z_{-i}\succ_{P_{i,-i}}x_{-i}\right\} =0.\label{eq:EqIndP}
\end{equation}
We may then simply set the mapping $\psi:PE\left(P\right)\to PE\left(P^{ind}\right)$
as the identity map and establish that $R\left(P_{i}^{ind},\psi\left(x\right)\right)\geq R\left(P_{i}^{ind},x\right)$
for all $x$ and $i$.

Clearly, ``$\triangleq$'' holds if and only if \eqref{eq:EqIndP}
is uniformly true across all $x\in PE\left(P^{ind}\right)$ and all
$i\in N$, but \eqref{eq:EqIndP} is not very explicit and in general
involves allocations that may not be Pareto efficient. Intuitively,
as the definitions of welfarism and paternalism do not impose any
restrictions on $P_{i,-i}$ between any two $x_{-i}$ and $y_{-i}$
that do not Pareto dominate each other, it should be ``easy'' for
\eqref{eq:EqIndP} to fail and for ``$\vartriangleright_{X}$''
to hold.

Theorem \ref{thm:IndMax}(b)(c) present two sufficient conditions
for ``$\vartriangleright_{X}$'' that formalizes this intuition
in a more precise manner. In particular, the conditions in Theorem
\ref{thm:IndMax}(c) are very explicit and arguably weak: $N\geq3$,
$\#\left(PE\left(P\right)\right)\geq2$ are almost trivial conditions
to check, and the focus on strict $\left(P_{i,-i}\right)$, though
not necessary, covers a representative class of social preferences.
The condition in Theorem \ref{thm:IndMax}(b) may seem less explicit,
but is in fact even weaker than the conditions in Theorem \ref{thm:IndMax}(c).

\section{\label{sec:Continuum}A Simple Example with Continuous Allocation
Space}

We now provide a simple example that illustrates how the key idea
underlying the partial order $\trianglerighteqslant$ proposed in
the previous sections for ordinal preference profiles on discrete
allocation space can be adapted to settings with cardinal preference
profiles on continuous allocation spaces.

Consider again the simple case with two individuals ``$1,2$'' and
two perfectly divisible goods ``$1,2$''. Let $x_{i}=\left(x_{i,1},x_{i,2}\right)$
denotes individual $i$'s consumption of goods $1,2$, respectively.
Suppose now that the society is endowed with 1 unit of each good,
so that the feasibility set is given by 
\[
X=\left\{ \left(x_{1},x_{2}\right)\in\R_{+}^{2}\times\R_{+}^{2}:\ \begin{array}{c}
x_{1,1}+x_{2,1}\leq1\\
x_{1,2}+x_{2,2}\leq1
\end{array}\right\} .
\]
Notice that now the cardinality of $X$ is uncountably infinite, but
the dimension is finite. It's now easier to work with cardinal preferences
encoded by a \emph{continuous} and \emph{weakly bivariate-increasing}
utility function $u_{i}:\R_{+}^{2}\to\R$.

With continuous $u_{i}$ and compact $X$, the levels of maximal and
minimal feasible utilities are well-defined:
\begin{align*}
\ol u_{i}=\sup_{x\in X}u_{i}\left(x\right) & ,\quad\ul u_{i}=\inf_{x\in X}u_{i}\left(x\right).
\end{align*}
Denote the set of feasible utilities as 
\begin{align*}
U & :=u\left(X\right).\\
U_{i} & :=u_{i}\left(X\right)=\left[\ul u_{i},\ol u_{i}\right]
\end{align*}

Given $X$ and $u$, we write $PE\left(u\right)$ to denote its Pareto
frontier. Here:

For each $x\in X$, and any utility function of individual $i$, we
define individual $i$'s ranking evaluation function $r_{i}\left(u_{i}\right):X\to\left[0,1\right]$
by
\begin{align*}
r_{i}\left(u_{i}\right)\left[x\right] & :=\frac{u_{i}\left(x\right)-u_{i}\left(x\right)}{\ol u_{i}-\ul u_{i}}
\end{align*}
with \textit{\textcolor{black}{``1'' corresponding to the most preferred
and ``0'' the least preferred}}\textcolor{black}{, }as in this example
we have
\[
-\infty<\ul u_{i}<\ol u_{i}<\infty.
\]

We now define a ranking evaluations of PE allocations by
\begin{align*}
rPE_{i}\left(u\right)\left[x\right] & :=\left\{ r_{i}\left(u_{i}\right)\left[x\right]:x\in PE\left(u\right)\right\} .
\end{align*}
and the Pareto frontier ranking evaluation profiles, $rPE\left(u,X\right):X\to U\subseteq\R^{2}$,
by
\[
rPE\left(u\right)\left[x\right]:=\left\{ r\left(u\right)\left[x\right]:x\in PE\left(u\right)\right\} .
\]

~

For simplicity, let's focus\textcolor{black}{{} on }\textit{\textcolor{black}{private
}}\textcolor{black}{utilities fi}rst for this example.

One ``preference profile'' $\ol u$ is given by
\[
\begin{cases}
\ol u_{1}\left(x_{1}\right)=x_{1,1}\\
\ol u_{2}\left(x_{2}\right)=x_{2,2}
\end{cases}
\]
which essentially makes $i$'s and $j$'s preferences ``orthogonal''
to each other, giving 
\begin{align*}
PE\left(\ol u\right) & =\left\{ \left(x_{1}=\left(\begin{array}{c}
1\\
0
\end{array}\right),x_{2}=\left(\begin{array}{c}
0\\
1
\end{array}\right)\right)\right\} 
\end{align*}
and thus
\[
rPE\left(\ol u\right)=\left\{ \left(1,1\right)\right\} .
\]

An alternative preference profile $\ul u$ is given by
\[
\ul u_{i}\left(x_{i}\right)=x_{i,1}+x_{i,2},\ \forall i\in\left\{ 1,2\right\} ,
\]
which essentially makes $i,j$'s utility function ``coincide perfectly'',
giving
\begin{align*}
PE\left(\ul u\right) & =\left\{ \left(x_{1}=\left(\begin{array}{c}
s\\
t
\end{array}\right),x_{2}=\left(\begin{array}{c}
1-s\\
1-t
\end{array}\right)\right):s,t\in\left[0,1\right]\right\} \\
U_{i} & =\left[0,2\right]
\end{align*}
and
\begin{align*}
rPE\left(\ul u\right)\left[x\right] & =\left\{ \left(\frac{\ul u_{1}\left(x\right)-0}{2-0},\frac{\ul u_{2}\left(x\right)-0}{2-0},\right):x\in PE\left(\ul u\right),x_{1}=\left(\begin{array}{c}
s\\
t
\end{array}\right)\right\} .\\
 & =\left\{ \left(\frac{s+t}{2},1-\frac{s+t}{2},\right):s,t\in\left[0,1\right]\right\} \\
 & =\left\{ \left(t,1-t\right):t\in\left[0,1\right]\right\} 
\end{align*}

The two preference profiles $\ol u$ and $\ul u$ can then be ``ordered''
based on the observation that
\[
rPE\left(\ol u\right)\geq rPE\left(\ul u\right)\left[x\right],\quad\forall x\in PE\left(\ul u\right).
\]

\section{\label{sec:Conclusion}Conclusion}

This paper proposes a theoretical framework under which preference
profiles can be meaningfully compared, without the need to aggregate
or trade off preferences across individuals. The current set of models
and results above are intended more as illustrations of the general
idea, and there are clearly many directions for further explorations. 

First, it would be interesting to generalize the illustrative example
in Section \eqref{sec:Continuum} so as to accommodate canonical continuous
allocation spaces and profiles of continuous utility functions, and
investigate how different configurations of social preferences can
be evaluated under the partial order. In particular, it would be theoretically
appealing if we can incorporate a ``numeraire'', on which every
individual's preference is perfectly aligned (e.g. everyone prefers
more money than less). Additionally, the framework can be further
enriched by introducing endowments into the setup, which helps establish
a selection from or a refinement of the Pareto frontier.

Second, it would be interesting to analyze various forms of social
preferences, potentially in a cardinal framework without the lexicographic
structure imposed between private preferences and social preferences
in Section \ref{subsec:Str_Social}. For example, a \emph{paternalistic}
social preference would be one such that an individual $i$ is happier
when other individuals obtain private allocations that are more favorable
according to $i$'s preference. A \emph{welfaristic} social preference
would be one such that an individual $i$ is happier when other individuals
obtain allocations that are more favorable according to other individuals'
own preferences.

Lastly, the current framework focuses on a fixed set of individuals.
Alternatively, one can ask whether the partial order can be adapted
to a setting with a large number (or a distribution) of individuals
with a corresponding distribution of preferences (or ``types'').
If the word ``ideology'' can be interpreted as a distribution over
preferences, at least in certain contexts, then the framework may
potentially serve as a formal basis for the comparison of different
ideologies.

\addcontentsline{toc}{section}{References}

\bibliographystyle{ecta}
\bibliography{PrefRank}

\appendix

\section{\label{sec:mainproof}Main Proofs}

\subsection{\label{subsec:pf_lem_AllPE}Proof of Lemma \ref{lem:AllPE}}
\begin{proof}
Suppose that $X\backslash PE\left(P,X\right)\neq\es$. Take any Pareto
dominated allocation $\ul x\in X\backslash PE\left(P,X\right)$, and
define
\begin{align*}
\ol{PE}_{\ul x}\left(P\right) & :=\left\{ y\in PE\left(P\right):R\left(P,y\right)\lneqq R\left(P,\ul x\right)\right\} ,\\
\ul{PE}_{\ul x}\left(P\right) & :=PE\left(P\right)\backslash\ol{PE}_{\ul x}\left(P\right),\\
NPE\left(P\right) & :=X\backslash PE\left(P\right).
\end{align*}
Clearly, $\ol{PE}_{\ul x}\left(P,X\right)\neq\es$ and
\[
X=\ol{PE}_{\ul x}\left(P\right)\cup\ul{PE}_{\ul x}\left(P\right)\cup NPE\left(P\right).
\]

Take any $x^{*}\in\ol{PE}_{\ul x}\left(P,X\right)$. there must exist
some individual $\ol i\in N$ such that
\begin{align*}
R\left(P_{\ol i},x^{*}\right) & <R\left(P_{\ol i},\ul x\right).
\end{align*}
Now, define
\[
\ol x:\in\arg\min_{y\in\ol{PE}_{\ul x}\left(P\right)}R\left(P_{\ol i},y\right)
\]
Define
\[
ID_{\ol x}:=\left\{ x\in X:x\sim_{P_{i}}\ol x\text{ for all }i\right\} 
\]
We must have
\[
R\left(P,\ol x\right)\leq R\left(P,\ul x\right),
\]
and moreover
\begin{align*}
R\left(P_{\ol i},\ol x\right) & \leq R\left(P_{\ol i},x^{*}\right)<R\left(P_{\ol i},\ul x\right).
\end{align*}

We now construct another  preference profile $P^{'}$ by switching
individual $\ol i$'s rankings of $\ul x$ and $\ol x$ only and keep
all other rankings unchanged:
\begin{align}
R\left(P_{\ol i}^{'},\ul x\right) & :=R\left(P_{\ol i},\ol x\right),\nonumber \\
R\left(P_{\ol i}^{'},\ol x\right) & :=R\left(P_{\ol i},\ul x\right)-\#ID_{\ol x}(+1),\nonumber \\
R\left(P_{\ol i}^{'},x\right) & :=R\left(P_{\ol i},x\right)-\#ID_{\ol x}\text{ for }x\text{ s.t. }\ol x\succ_{P_{\ol i}}x\succ_{P_{\ol i}}\ul x\\
R\left(P_{\ol i}^{'},x\right) & :=R\left(P_{\ol i},x\right),\ \forall x\in X\backslash\left\{ \ul x,\ol x\right\} ,\nonumber \\
R\left(P_{i}^{'},x\right) & :=R\left(P_{i},x\right),\ \forall i\in N\backslash\left\{ \ol i\right\} ,\forall x\in X.\label{eq:def_newpref-1}
\end{align}
Notice that this construction ensures that
\begin{align}
R\left(P^{'},\ul x\right) & \lneqq R\left(P,\ul x\right),\label{eq:x_l_dec-1}\\
R\left(P^{'},\ol x\right) & \gneqq R\left(P,\ol x\right),\label{eq:x_u_inc-1}\\
R\left(P^{'},\ul x\right) & \geq R\left(P,\ol x\right).\label{eq:x_lu-1}
\end{align}

We now show that $\ul x\in PE\left(P^{'}\right)$. First, notice that
$\ul x$ is not Pareto dominated by any allocation $x\in\ol{PE}_{\ul x}\left(P\right)$
under $P^{'}$: by construction, 
\begin{align*}
R\left(P_{\ol i}^{'},\ul x\right) & =R\left(P_{\ol i},\ol x\right)\\
 & \leq R\left(P_{\ol i},x\right),\ \forall x\in\ol{PE}_{\ul x}\left(P\right)\backslash\left\{ \ol x\right\} ,\\
 & =R\left(P_{\ol i}^{'},x\right),\ \forall x\in\ol{PE}_{\ul x}\left(P\right)\backslash\left\{ \ol x\right\} .
\end{align*}
Moreover, $\ul x$ is not dominated by any allocation $x\in\left(X\backslash\ol{PE}_{\ul x}\left(P\right)\right)\backslash\left\{ \ul x\right\} $
under $P^{'}$: if $x\in X\backslash\ol{PE}_{\ul x}\left(P\right)$
and $x\neq\ul x$, then either (i) there must exist some $i\in N$
such that $R\left(P_{i},x\right)>R\left(P_{i},\ul x\right),$or (ii):
$R\left(P,x\right)=R\left(P,\ul x\right).$ For case (i), by \eqref{eq:x_l_dec-1}
we must have 
\begin{align*}
R\left(P_{i}^{'},x\right) & =R\left(P_{i},x\right)>R\left(P_{i},\ul x\right)\geq R\left(P_{i}^{'},\ul x\right)
\end{align*}
For case (ii), by \eqref{eq:x_l_dec-1} we must have
\begin{align*}
R\left(P^{'},x\right) & =R\left(P,x\right)=R\left(P,\ul x\right)\geq R\left(P,\ul x\right).
\end{align*}
In either case, $x$ cannot Pareto dominate $\ul x$.

Next we show that $PE\left(P\right)\backslash\left\{ \ol x\right\} \subseteq PE\left(P^{'}\right)$.
For any $x\in PE\left(P\right)\backslash\left\{ \ol x\right\} $,
it is not Pareto dominated by any $x^{'}\in X\backslash\left\{ \ul x\right\} $
under $P$, so it must not be Pareto dominated by $x$ under $P^{'}$,
as the ranking profile of $x$ stay unchanged while the ranking profile
of $x^{'}$ weakly worsens:
\[
R\left(P^{'},x^{'}\right)\geq R\left(P,x^{'}\right),\ \forall x^{'}\in X\backslash\left\{ \ul x\right\} .
\]
In particular, recall that by \eqref{eq:x_u_inc-1} the ranking profile
of $\ol x$ strictly worsens from $P$ to $P^{'}$. Moreover, $x$
cannot be Pareto dominated by $\ul x$ either. We consider three cases
separately. 

Case (i): $x\in\ul{PE}_{\ul x}\left(P\right)$, i.e., $x$ is Pareto
efficient but $x$ does not Pareto dominate $\ul x$. In this case,
\begin{equation}
R\left(P,x\right)\neq R\left(P,\ul x\right),\label{eq:notalleq}
\end{equation}
otherwise $x$ would be Pareto dominated by $\ol x$ under $P$. Then,
given $\ul x$ is not Pareto dominated by $x$ under $P$ and \eqref{eq:notalleq},
there must exist some $i\in N$ such that
\[
R\left(P_{i},x\right)>R\left(P_{i},\ul x\right)\geq R\left(P_{i},\ol x\right).
\]
Now, given that $x$ is not Pareto dominated by $\ol x$ under $P$,
there must exist some $j\in N$ such that
\[
R\left(P_{j},x\right)<R\left(P_{j},\ol x\right).
\]
As $x\notin\left\{ \ul x,\ol x\right\} $, $R\left(P^{'},x\right)=R\left(P,x\right)$
by \eqref{eq:def_newpref-1}, we have, if $j=\ol i$,
\[
R\left(P_{j}^{'},x\right)<R\left(P_{j},\ol x\right)=R\left(P_{j}^{'},\ul x\right),
\]
and if $j\neq\ol i$,
\[
R\left(P_{j}^{'},x\right)<R\left(P_{j},\ol x\right)\leq R\left(P_{j},\ul x\right)\leq R\left(P_{j}^{'},\ul x\right),
\]
implying that $x$ cannot be Pareto dominated by $\ul x$ under $P^{'}$.

Now, for $x\in\ol{PE}_{\ul x}\left(P\right)\backslash\left\{ \ol x\right\} $,
as $x$ and $\ol x$ do not Pareto dominate each other, we have two
additional cases.

Case (ii): $x\in\ol{PE}_{\ul x}\left(P\right)\backslash\left\{ \ol x\right\} $
and there exists some $j\in N$ such that 
\[
R\left(P_{j},x\right)<R\left(P_{j},\ol x\right).
\]
Then we can apply the same arguments as in Case (i) and deduce that
\[
R\left(P_{j}^{'},x\right)<R\left(P_{j}^{'},\ul x\right),
\]
implying that $x$ cannot be Pareto dominated by $\ul x$ under $P^{'}$.

Case (iii): $x\in\ol{PE}_{\ul x}\left(P\right)\backslash\left\{ \ol x\right\} $
and $R\left(P,x\right)=R\left(P,\ol x\right)$, i.e., all individuals
are indifferent between $x$ and $\ol x$. Then, as $x\notin\left\{ \ol x,\ul x\right\} $,
\begin{align*}
R\left(P^{'},x\right) & =R\left(P,x\right)=R\left(P,\ol x\right)\\
 & \begin{cases}
=R\left(P_{\ol i}^{'},\ul x\right),\\
\geq R\left(P_{j},\ul x\right), & j\neq\ol i
\end{cases}\\
 & \geq R\left(P^{'},\ul x\right),
\end{align*}
implying that $x$ cannot be Pareto dominated by $\ul x$ under $P^{'}$.

In summary of the above, we have shown that 
\[
PE\left(\ul P\right)\cup\left\{ \ul x\right\} \backslash\left\{ \ol x\right\} \subseteq PE\left(P^{'}\right)
\]
which in particular implies that $\#\left(PE\left(\ul P\right)\right)\leq\#\left(PE\left(P^{'}\right)\right).$

Now consider the set
\begin{align*}
\D:=PE\left(P^{'}\right)\backslash\left(PE\left(P\right)\cup\left\{ \ul x\right\} \right),
\end{align*}
which may or may not be empty. In particular, notice that $\ol x\notin\D$
as $\ol x\in PE\left(P\right)$.

Suppose that $\D\neq\es$. Take any $x\in\D$. Clearly, $x\in NPE\left(P\right)$.
We show that $x$ must be Pareto dominated by $\ol x$ under $P$,
i.e.,
\begin{align*}
R\left(P,x\right) & \gneqq R\left(P,\ol x\right).
\end{align*}
To see this, notice that the ranking profile of $x$ must have remained
unchanged from $P$ to $P^{'}$, while only the ranking profile of
$\ol x$ in $X$ has worsened from $P$ to $P^{'}$, i.e.
\begin{align*}
R\left(P^{'},x\right) & =R\left(P,x\right),\\
R\left(P^{'},\ol x\right) & \gneqq R\left(P,\ol x\right),\\
R\left(P^{'},y\right) & \leq R\left(P,y\right),\ \forall y\in X\backslash\left\{ \ol x\right\} .
\end{align*}
As a result, if $x$ is Pareto dominated by some $y\in X\backslash\ol x$
under $P$, $x$ must still be Pareto dominated by $y$ under $P^{'}$.
Hence, by the fact that $x$ is Pareto dominated under $P$ but becomes
Pareto efficient under $P^{'}$, it must be the case that $x$ is
Pareto dominated by $\ol x$ under $P$. Then, we must have
\begin{equation}
R\left(P^{'},x\right)=R\left(P,x\right)\gneqq R\left(P,\ol x\right).\label{eq:x_new_inc-1}
\end{equation}

Now, observe that
\[
PE\left(P^{'}\right)=\left[PE\left(P\right)\backslash\left\{ \ol x\right\} \right]\cup\left\{ \ul x\right\} \cup\D\cup\left[PE\left(P^{'}\right)\cap\left\{ \ol x\right\} \right],
\]
where the last term $PE\left(P^{'}\right)\cap\left\{ \ol x\right\} $
may either be nonempty or empty, depending on whether $\ol x\in PE\left(P^{'}\right)$. 

We now define the mapping $\psi:PE\left(P^{'},X\right)\to PE\left(P,X\right)$
by
\begin{align*}
\psi\left(x\right) & :=\begin{cases}
x, & x\in PE\left(P\right)\backslash\left\{ \ol x\right\} ,\\
\ol x, & x\in\left\{ \ul x\right\} ,\\
\ol x, & x\in\D,\\
\ol x, & x\in PE\left(P^{'}\right)\cap\left\{ \ol x\right\} ,
\end{cases}\\
 & =\begin{cases}
x, & x\in PE\left(P\right)\backslash\left\{ \ol x\right\} ,\\
\ol x, & x\in PE\left(P^{'}\right)\backslash\left(PE\left(P\right)\backslash\left\{ \ol x\right\} \right),
\end{cases}
\end{align*}
which is clearly an onto mapping:
\[
\psi\left(PE\left(P^{'}\right)\right)=PE\left(P\right).
\]
Moreover, for every $x\in PE\left(P^{'}\right)$, we have
\begin{align*}
R\left(P,\psi\left(x\right)\right) & =\begin{cases}
R\left(P,x\right), & x\in PE\left(P\right)\backslash\left\{ \ol x\right\} ,\\
R\left(P,\ol x\right) & x\in\left\{ \ul x\right\} ,\\
R\left(P,\ol x\right) & x\in\D,\\
R\left(P,\ol x\right) & x\in PE\left(P^{'}\right)\cap\left\{ \ol x\right\} ,
\end{cases}\\
 & \sim\begin{cases}
=R\left(P^{'},x\right), & x\in PE\left(P\right)\backslash\left\{ \ol x\right\} ,\quad\text{by \eqref{eq:def_newpref-1}},\\
\leq R\left(P^{'},\ul x\right), & x\in\left\{ \ul x\right\} ,\quad\text{by }\eqref{eq:x_lu-1},\\
<R\left(P^{'},x\right), & x\in\D,\quad\text{by }\eqref{eq:x_new_inc-1},\\
\lneqq R\left(P^{'},\ol x\right) & x\in PE\left(P^{'}\right)\cap\left\{ \ol x\right\} ,\quad\text{by }\eqref{eq:x_u_inc-1},
\end{cases}\\
 & =R\left(P^{'},x\right).
\end{align*}

In summary, we have established the existence of an onto mapping $\psi:PE\left(P^{'}\right)\to PE\left(P\right)$
such that $R\left(P,\psi\left(x\right)\right)\leq R\left(P^{'},x\right)$
for every $x\in PE\left(P^{'}\right)$ with at least one strict inequality.
Hence, $P\succ_{X}P^{'}$, so $P$ is not a minimal element.
\end{proof}

\subsection{\label{subsec:pf_lem_no22}Proof of Lemma \ref{lem:No22indiff}}
\begin{proof}
Suppose that there exist two distinct individuals $i,j\in N$ and
a Pareto efficient allocation $x\in PE\left(P\right)$ such that $x\sim_{P_{i}}y,\:x\sim_{P_{j}}y$
for some $y\in X\backslash\left\{ x\right\} $. We construct another
 preference profile $P^{'}$ in the following way. 

\noindent \textbf{Construction of $P^{'}$}

We keep the preferences of all individuals other than $i,j$ unchanged,
i.e., $P_{k}^{'}=P_{k}$ for all $k\in N\backslash\left\{ i,j\right\} $. 

For individual $i$, we construct $P_{i}^{'}$ by perturbing $P_{i}$
in the following way. Define 
\[
{\cal I}_{x}\left(P_{i}\right):=\left\{ z\in X\backslash\left\{ x\right\} :z\sim_{P_{i}}x\right\} ,
\]
First, we set
\[
x\succ_{P_{i}^{'}}y.
\]
Second, if there exists any $z\in{\cal I}_{x}\left(P_{i}\right)\backslash\left\{ y\right\} $,
we set
\[
z\sim_{P_{i}^{'}}y.
\]
In other words, the allocation $x$ is preferred under $P_{i}^{'}$
to any other allocation $z$ that $i$ found indifferent with $x$
under $P_{i}$, i.e.,
\[
x\succ_{P_{i}^{'}}z,\quad\forall z\in{\cal I}_{x}\left(P_{i}\right).
\]
Third, we keep all other pairwise preferences in $P_{i}$ unchanged.

For individual $j$, we construct $P_{j}^{'}$ by perturbing $P_{j}$
in the following way. Now, write
\[
{\cal I}_{x}\left(P_{j}\right):=\left\{ z\in X\backslash\left\{ x\right\} :z\sim_{P_{j}}x\right\} ,
\]
First, we set
\[
y\succ_{P_{j}^{'}}x.
\]
Second, if there exists any $z\in{\cal I}_{x}\left(P_{j}\right)\backslash\left\{ y\right\} $,
we set
\[
z\sim_{P_{j}^{'}}y.
\]
In other words, any other allocation $z$ that $j$ found indifferent
with $x$ under $P_{j}$, is now preferred under $P_{j}^{'}$ to allocation
$x$:
\[
z\succ_{P_{j}^{'}}x,\quad\forall z\in{\cal I}_{x}\left(P_{j}\right).
\]

Note that, for any two allocations $u,v\in X$, individual preferences
between $u$ and $v$ under $P$ and $P^{'}$ can be different only
if $u=x$ and $v\in{\cal I}_{x}\left(P_{i}\right)\cup{\cal I}_{x}\left(P_{j}\right)$
or vice versa.

\noindent \textbf{Characterization of $PE\left(P^{'}\right)$}

\noindent We claim that 
\[
PE\left(P^{'}\right)=PE\left(P\right)\cup\left\{ z\in\text{\ensuremath{{\cal I}_{x}\left(P_{j}\right)}: \ensuremath{z} is Pareto dominated under \ensuremath{P} by }x\text{ \text{and only }}x\right\} .
\]
We prove this claim by considering the following cases:
\begin{enumerate}
\item We show 
\[
x\in PE\left(P^{'}\right).
\]
Recall that $x\in PE\left(P\right)$ by supposition, and consider
the following cases:
\begin{enumerate}
\item For any $z\in{\cal I}_{x}\left(P_{i}\right)$, i.e., $z\sim_{P_{i}}x$,
we know that $x\succ_{P_{i}^{'}}z$, so $x$ cannot be Pareto dominated
by $z$.
\item For any $z\in{\cal I}_{x}\left(P_{j}\right)\backslash{\cal I}_{x}\left(P_{i}\right)$,
i.e., $z\sim_{P_{j}}x$ but $z\nsim_{P_{i}}x,$, we consider the following
two possibilities. If $x\succ_{P_{i}}z$, then this pairwise comparison
remain unchanged from $P$ to $P^{'}$, so we have $x\succ_{P_{i}^{'}}z$
by the construction of $P^{'}$ and thus $x$ cannot be Pareto dominated
by $z$. If otherwise $z\succ_{P_{i}}x$, then we can deduce from
$x\in PE\left(P\right)$ that there must exist some individual $k\notin\left\{ i,j\right\} $
with $x\succ_{P_{k}}z$. Again, since we did not perturb the preference
of any $k\notin\left\{ i,j\right\} $in the construction of $P^{'}$,
we have $x\succ_{P_{k}^{'}}z$, so $x$ cannot be Pareto dominated
by $z$.
\item For any $z\notin{\cal I}_{x}\left(P_{i}\right)\cup{\cal I}_{x}\left(P_{j}\right)\cup\left\{ x\right\} $,
i.e. $z\nsim_{P_{i}}x$ and $z\nsim_{P_{j}}x$, the pairwise comparison
between $x$ and $z$ stays completely unchanged from $P$ to $P^{'}$.
Hence, given that $x$ was not Pareto dominated by $z$ under $P$,
it remains under $P^{'}$.
\end{enumerate}
Combining the three collectively exhaustive cases above, we conclude
that $x$ is not Pareto dominated by any $x\in X\backslash\left\{ x\right\} $
under $P^{'}$, or equivalently, $x\in PE\left(P^{'}\right)$.
\item We show 
\[
PE\left(P\right)\cap{\cal I}_{x}\left(P_{i}\right)\cap{\cal I}_{x}\left(P_{j}\right)\subseteq PE\left(P^{'}\right).
\]
To see this, consider any $z\in PE\left(P\right)\cap{\cal I}_{x}\left(P_{i}\right)\cap{\cal I}_{x}\left(P_{j}\right)$. 
\begin{enumerate}
\item For $x$, by the construction of $P'$ we have $x\succ_{P_{i}^{'}}z$
and $z\succ_{P_{j}^{'}}x$, so $z$ cannot be dominated by $x$ under
$P^{'}$.
\item For any $w\in{\cal I}_{x}\left(P_{i}\right)\cap{\cal I}_{x}\left(P_{j}\right)$,
we have $z\sim_{P_{i}^{'}}w$ and $z\sim_{P_{j}^{'}}w$ by the construction
of $P^{'}$ and the preferences between $z$ and $w$ of any other
individuals $k\notin\left\{ i,j\right\} $ stay unchanged from $P$
to $P^{'}$. Hence, if $z$ is not Pareto dominated by $x$ under
$P$, it remains so under $P^{'}$.
\item For any $w\in{\cal I}_{x}\left(P_{i}\right)\backslash{\cal I}_{x}\left(P_{j}\right)$,
we consider two possibilities given $w\notin{\cal I}_{x}\left(P_{j}\right)$.
If $z\succ_{P_{j}}w$, then $z\succ_{P_{j}^{'}}w$ and thus $z$ is
not Pareto dominated by $w$ under $P^{'}$. Otherwise if $w\succ_{P_{j}}z$,
then by $z\in PE\left(P\right)\cap{\cal I}_{x}\left(P_{i}\right)$
there must exist some $k\notin\left\{ i,j\right\} $ such that $z\succ_{P_{k}}w$.
Then $P_{k}^{'}=P_{k}$ by construction and $z\succ_{P_{k}^{'}}x$,
so again $z$ is not Pareto dominated by $w$ under $P^{'}$.
\item For any $w\in{\cal I}_{x}\left(P_{j}\right)\backslash{\cal I}_{x}\left(P_{i}\right)$,
the arguments above in 2(c) apply with $i$ in place of $j$.
\item For any $w\in X\backslash\left(\left\{ x\right\} \cup{\cal I}_{x}\left(P_{i}\right)\cup{\cal I}_{x}\left(P_{j}\right)\right)$,
the preference between $z$ and $w$ of any individual remains unchanged.
Hence, if $z$ is not Pareto dominated by $x$ under $P$, it remains
so under $P^{'}$.
\end{enumerate}
Combing the five collectively exhaustive cases above, we conclude
that $z\in PE\left(P^{'}\right)$.
\item We show
\[
PE\left(P\right)\cap{\cal I}_{x}\left(P_{i}\right)\backslash{\cal I}_{x}\left(P_{j}\right)\subseteq PE\left(P^{'}\right).
\]
To see this, consider any $z\in PE\left(P\right)\cap{\cal I}_{x}\left(P_{i}\right)\backslash{\cal I}_{x}\left(P_{j}\right)$.
\begin{enumerate}
\item For $x$, we consider two possibilities given that $z\notin{\cal I}_{x}\left(P_{j}\right)$.
If $z\succ_{P_{j}}x$, then we have $z\succ_{P_{j}^{'}}x$ and thus
$z$ is not Pareto dominated by $x$ under $P^{'}$. Otherwise if
$x\succ_{P_{j}}z$, then since $z\in PE\left(P\right)\cap{\cal I}_{x}\left(P_{i}\right)$
there must exist some $k\notin\left\{ i,j\right\} $ such that $z\succ_{P_{k}}x$.
Then $P_{k}^{'}=P_{k}$ by construction and $z\succ_{P_{k}^{'}}x$,
so again $z$ is not Pareto dominated by $x$ under $P^{'}$.
\item For $w\in X\backslash\left\{ x\right\} $, no individual's preference
between $z$ and $w$ has been changed from $P$ to $P^{'}$. (In
particular, if $w\sim_{P_{i}}z\sim_{P_{i}}x$, then under $P^{'}$
we have $x\succ_{P_{i}^{'}}w\sim_{P_{i}^{'}}z$, so the preference
between $z$ and $w$ stays unchanged from $P$ to $P^{'}$.\footnote{This can be seen more clearly by considering the following three sub-cases
separately:
\begin{enumerate}
\item For $w\in{\cal I}_{x}\left(P_{i}\right)\backslash{\cal I}_{x}\left(P_{j}\right)$,
we have $z\sim_{P_{i}^{'}}w$ and no other individual's preference
between $z$ and $w$ is changed from $P$ to $P^{'}$. Hence, given
that $z$ is not Pareto dominated by $w$ under $P$, it remains so
under $P^{'}$.
\item For $w\in{\cal I}_{x}\left(P_{j}\right)\backslash{\cal I}_{x}\left(P_{i}\right)$,
we know that both $i$'s and $j$'s preferences betweeen $z$ and
$w$ stay unchanged from $P$ to $P^{'}$. Hence, given that $z$
is not Pareto dominated by $w$ under $P$, it remains so under $P^{'}$.
\item For any $w\in X\backslash\left(\left\{ x\right\} \cup{\cal I}_{x}\left(P_{i}\right)\cup{\cal I}_{x}\left(P_{j}\right)\right)$,
we note that the comparison between $w$ and $x$ does not change
at all from $P$ to $P^{'}$. Hence, given that $z$ was not Pareto
dominated by $w$ under $P$, it remains so under $P^{'}$.
\end{enumerate}
}
\end{enumerate}
\item Similarly to case 3, we have
\[
PE\left(P\right)\cap{\cal I}_{x}\left(P_{j}\right)\backslash{\cal I}_{x}\left(P_{i}\right)\subseteq PE\left(P^{'}\right).
\]
\item We have
\[
PE\left(P\right)\backslash\left(\left\{ x\right\} \cup{\cal I}_{x}\left(P_{i}\right)\cup{\cal I}_{x}\left(P_{j}\right)\right)\subseteq PE\left(P^{'}\right),
\]
since every individual's preference between $z\in PE\left(P\right)\backslash\left(\left\{ x\right\} \cup{\cal I}_{x}\left(P_{i}\right)\cup{\cal I}_{x}\left(P_{j}\right)\right)$
and any $w\in X$ stays unchanged from $P$ to $P^{'}$.\textbackslash{}
\end{enumerate}
Combining Points 1-5 above, we have
\begin{equation}
PE\left(P\right)\subseteq PE\left(P^{'}\right).\label{eq:PE_inc}
\end{equation}
Now we analyze what happens to every $z\notin PE\left(P\right)$ after
the change from $P$ to $P^{'}$. Specifically, we separately consider
the following possibilities: \setcounter{enumi}{5}
\begin{enumerate}
\item If $z\in X\backslash\left(PE\left(P\right)\cup{\cal I}_{x}\left(P_{j}\right)\right)$,
we can show that
\[
z\notin PE\left(P^{'}\right).
\]
To see this, notice that $z\notin PE\left(P\right)$ implies that
$z$ is Pareto dominated by some allocation $w\in X$. If $w\neq x$,
then $z$ will remain Pareto dominated by $w$, since everyone's preference
between $z$ and $w$ stays unchanged from $P$ to $P^{'}$. If $w=x$,
then from $P$ to $P^{'}$ an individual's preference of $z$ relative
to $x$ can strictly improve only for individual $j$ and only if
$z\in{\cal I}_{x}\left(P_{j}\right)$; however, since $z\notin{\cal I}_{x}\left(P_{j}\right)$,
we conclude that $z$ remains Pareto dominated by $x$ under $P^{'}$.
\item If $z\in\left(X\backslash PE\left(P\right)\right)\cap{\cal I}_{x}\left(P_{j}\right)$
and $z$ is Pareto dominated by some $w\in X\backslash\left\{ x\right\} $
under $P$, we can again show that
\[
z\notin PE\left(P^{'}\right).
\]
To see this, notice that every individual's preference between $z$
and $w$ stays unchanged from $P$ to $P^{'}$. Hence, given that
$z$ is Pareto dominated by $w$ under $P$, it remains so under $P^{'}$.
\item If $z\in\left(X\backslash PE\left(P\right)\right)\cap{\cal I}_{x}\left(P_{j}\right)$
and $z$ is Pareto dominated by $x$ and only $x$ under $P$, we
now show that 
\[
z\in PE\left(P^{'}\right).
\]
To see this, notice that $z$ cannot be Pareto dominated by $x$ under
$P^{'}$, since we have set
\[
z\succ_{P_{j}^{'}}x
\]
given that $z\in{\cal I}_{x}\left(P_{j}\right)$. In the meanwhile,
since $z$ is not Pareto dominated by any $w\in X\backslash\left\{ x\right\} $
under $P$, it remains so under $P^{'}$, given that every individual's
preference between $z$ and $w$ stays unchanged from $P$ to $P^{'}$.
\end{enumerate}
Combining \eqref{eq:PE_inc} with Points 6-8 above, we deduce that
\[
PE\left(P^{'}\right)=PE\left(P\right)\cup\left\{ z\in\text{\ensuremath{{\cal I}_{x}\left(P_{j}\right)}: \ensuremath{z} is Pareto dominated under \ensuremath{P} by }x\text{ \text{and only }}x\right\} .
\]

\subsubsection*{Construction of Mapping $\phi$ and Proof of Ordering}

We now construct the mapping $\phi:PE\left(P^{'}\right)\to PE\left(P\right)$
by 
\[
\phi\left(z\right)=\begin{cases}
z, & z\in PE\left(P\right),\\
x, & z\in PE\left(P^{'}\right)\backslash PE\left(P\right).
\end{cases}
\]
and prove that 
\[
R\left(P^{'},z\right)\geq R\left(P,\phi\left(z\right)\right)\quad\forall z\in PE\left(P^{'}\right),
\]
with at least one strictly inequality:
\begin{enumerate}
\item For $z=x$, we have $\phi\left(x\right)=x$ and
\begin{align*}
R\left(P_{i}^{'},x\right) & =R\left(P_{i},x\right),\\
R\left(P_{j}^{'},x\right) & =R\left(P_{j},x\right)+\#\left(\ensuremath{{\cal I}_{x}\left(P_{j}\right)}\right)\\
 & \geq R\left(P_{j},x\right)+1>R\left(P_{j},x\right),\\
R\left(P_{k}^{'},x\right) & =R\left(P_{k},x\right),\quad\forall k\in N\backslash\left\{ i,j\right\} .
\end{align*}
since we know $y\in\ensuremath{{\cal I}_{x}\left(P_{j}\right)}$ and
thus $\#\left(\ensuremath{{\cal I}_{x}\left(P_{j}\right)}\right)\geq1$
and thus the inequality $R\left(P_{j}^{'},x\right)>R\left(P_{j},x\right)$
must be strict.
\item For $z\in PE\left(P\right)\cap{\cal I}_{x}\left(P_{i}\right)$, we
have $\phi\left(z\right)=z$ and
\begin{align*}
R\left(P_{i}^{'},z\right) & =R\left(P_{i}^{'},x\right)+1\\
 & =R\left(P_{i},x\right)+1>R\left(P_{i},x\right)=R\left(P_{i},z\right),\\
R\left(P_{j}^{'},z\right) & =R\left(P_{j},z\right),\\
R\left(P_{k}^{'},x\right) & =R\left(P_{k},x\right),\quad\forall k\in N\backslash\left\{ i,j\right\} .
\end{align*}
\item For $z\in PE\left(P\right)\backslash\left[\left\{ x\right\} \cup{\cal I}_{x}\left(P_{i}\right)\right]$,
we have $\phi\left(z\right)=z$ and
\begin{align*}
R\left(P_{i}^{'},z\right) & =R\left(P_{i},z\right),\\
R\left(P_{j}^{'},z\right) & =R\left(P_{j},z\right),\\
R\left(P_{k}^{'},x\right) & =R\left(P_{k},x\right),\quad\forall k\in N\backslash\left\{ i,j\right\} .
\end{align*}
\item For $z\in PE\left(P^{'}\right)\backslash PE\left(P\right)$, we know
that $z\in{\cal I}_{x}\left(P_{j}\right)$ and $z$ is Pareto dominated
by $x$ and only $x$ under $P$. Hence, recalling that $\phi\left(z\right)=x$,
we have
\begin{align*}
R\left(P_{i}^{'},z\right) & \geq R\left(P_{i},z\right)\geq R\left(P_{i},x\right),\\
R\left(P_{j}^{'},z\right) & =R\left(P_{j},z\right)\geq R\left(P_{j},x\right),\\
R\left(P_{k}^{'},z\right) & =R\left(P_{k},z\right)\geq R\left(P_{k},x\right),\quad\forall k\in N\backslash\left\{ i,j\right\} .
\end{align*}
\end{enumerate}
Hence, $P$ is not minimal.
\end{proof}

\subsection{\label{subsec:pf_lem_strict}Proof of Lemma \ref{lem:AllStrict}}
\begin{proof}
Suppose that there exists a Pareto efficient allocation $x\in PE\left(P\right)$
such that $x\sim_{P_{i}}y$ for some $i$ and some $y\neq x$. 

If there exists another individual $j\neq i$ such that $x\sim_{P_{j}}y$,
then the condition of Lemma \ref{lem:No22indiff} is satisfied, so
the proof and conclusion of Lemma \ref{lem:No22indiff} apply.

Hence, here we focus on the remaining case, in which
\begin{equation}
x\nsim_{P_{j}}y,\quad\forall j\neq i.\label{eq:no_j_indff}
\end{equation}

We construct another  preference profile $P^{'}$ by setting 
\[
x\succ_{P_{i}^{'}}y\sim_{P_{i}^{'}}z,\quad\forall z\in{\cal I}_{x}\left(P_{i}\right)\backslash\left\{ y\right\} ,
\]
where 
\[
{\cal I}_{x}\left(P_{i}\right):=\left\{ z\in X\backslash\left\{ x\right\} :z\sim_{P_{i}}x\right\} .
\]
We keep all other pairwise preference relations completely unchanged
from $P$ to $P^{'}$.

We claim that $PE\left(P^{'}\right)=PE\left(P\right)$.

We first show that $PE\left(P\right)\subseteq PE\left(P^{'}\right)$
by considering the following three cases separately:
\begin{enumerate}
\item First, $x$ is clearly not Pareto dominated under $P^{'}$ by construction.
\item Second, we show that any $z\in PE\left(P\right)\cap{\cal I}_{x}\left(P_{i}\right)$
is not Pareto dominated under $P^{'}$. To see this, notice first
that $z$ is not Pareto dominated by $x$ under $P$. By \eqref{eq:no_j_indff},
there is no other individual $j$ such that $z\sim_{P_{j}}x.$ Hence,
there must exist two individuals $j,k\in N\backslash\left\{ i\right\} $
such that
\[
x\succ_{P_{j}}z,\quad\text{and}\quad z\succ_{P_{k}}x.
\]
Since $j,k$'s preferences stay unchanged from $P$ to $P^{'},$it
follows that
\[
x\succ_{P_{j}^{'}}z,\quad\text{and}\quad z\succ_{P_{k}^{'}}x,
\]
implying that $z$ is not Pareto dominated by $x$. As the preference
relation between $z$ and any other $w\in X\backslash\left\{ x,z\right\} $
is unchanged from $P$ to $P^{'}$, we conclude that $z\in PE\left(P^{'},X\right)$.
\item Third, any $z\in X\backslash\left[\left\{ x\right\} \cup{\cal I}_{x}\left(P_{i}\right)\right]$
cannot be Pareto dominated under $P^{'}$, as the preferences between
$z$ and any other $w\in X\backslash\left\{ z\right\} $ stays unchanged
from $P$ to $P^{'}$.
\end{enumerate}
Combining Points 1-3, we deduce that $PE\left(P\right)\subseteq PE\left(P^{'}\right)$.

We now show that $X\backslash PE\left(P\right)\subseteq X\backslash PE\left(P^{'}\right).$

For any $z\in X\backslash PE\left(P\right)$, it must be Pareto dominated
by some $w\in PE\left(P\right)$. If $w=x$, then $z$ remains Pareto
dominated by $x$ under $P^{'}$, since from $P$ to $P^{'}$ the
ranking vector of $x$ has weakly improved while the ranking vector
of $z$ cannot increase. If $w\neq x$, then every individual's preference
between $z$ and $w$ stays unchanged from $P$ to $P^{'}$, so $z$
must remain Pareto dominated by $w$ under $P^{'}$.

Hence, we conclude that $PE\left(P^{'}\right)=PE\left(P\right).$

~

Setting $\psi:PE\left(P^{'},X\right)\to PE\left(P,X\right)$ as the
identity mapping $\psi\left(z\right)=z$ for all $z\in PE\left(P,X\right)$,
we have
\begin{align*}
R\left(P_{i}^{'},x\right) & =R\left(P_{i},x\right),\\
R\left(P_{i}^{'},z\right) & =R\left(P_{i},z\right)+1>R\left(P_{i},z\right),\quad\forall z\in{\cal I}_{x}\left(P_{i}\right),\\
R\left(P_{i}^{'},z\right) & =R\left(P_{i},z\right),\quad\forall z\in X\backslash\left(\left\{ x\right\} \cup{\cal I}_{x}\left(P_{i}\right)\right),\\
R\left(P_{j}^{'},z\right) & =R\left(P_{j},z\right)\quad\forall j\neq i,\forall z\in X,
\end{align*}
and thus $R\left(P^{'},\psi\left(z\right)\right)\geq R\left(P,z\right)$
for all $z\in PE\left(P,X\right)$ with at least one strict inequality.
\end{proof}

\subsection{\label{subsec:pf_thm_strict}Proof of Theorem \ref{thm:discrete_gen}}
\begin{proof}
~
\begin{itemize}
\item[(a)]  $R\left(\ol P,x^{*}\right)={\bf 1}_{N}\leq R\left(P,x\right)$ for
all $x\in X$ and all possible preference profile $P$.
\item[(b)]  The ``only if'' part immediate from Lemma \ref{lem:AllPE} and
Lemma \ref{lem:AllStrict}. Here we prove the ``if'' part:

Given any other preference profile $P^{'}$, we show that it cannot
be the case that $P\succ_{X}P^{'}.$ We prove by contradiction and
suppose that $P\succ_{X}P^{'}.$ First, notice that by the supposition
that $PE\left(\ul P\right)=X$, for there to exist an onto mapping
$\psi:PE\left(P^{'}\right)\to PE\left(\ul P\right)$, it must be the
case that 
\[
PE\left(P^{'}\right)=PE\left(\ul P\right)=X,
\]
 and $\psi$ is a permutation on $X$. Moreover, by the supposition
we must have
\[
R\left(P^{'},x\right)\geq R\left(\ul P,\psi\left(x\right)\right),\quad\text{for all }x\in X.
\]
and 
\[
R\left(P^{'},x\right)\gneqq R\left(\ul P,\psi\left(x\right)\right),\quad\text{for some }x\in X.
\]
Then, by the summing over all individuals $i\in N$ and all $x\in X$,
we have
\begin{equation}
\sum_{i\in N}\sum_{x\in X}R\left(P_{i}^{'},x\right)>\sum_{i\in N}\sum_{x\in X}R\left(\ul P_{i},\psi\left(x\right)\right)\label{eq:ineq_sumrank}
\end{equation}
By the supposition that $\ul P$ is strict, each individual's ranking
of $x\in X$ must be a permutation of $\left(1,...,M\right),$ while
under a general $P^{'}$, each ranking vector must be weakly dominated
by $\left(1,...,M\right),$ so we have
\[
\sum_{i\in N}\sum_{x\in X}R\left(P_{i}^{'},x\right)\leq\frac{1}{2}NM\left(M+1\right)=\sum_{i\in N}\sum_{x\in X}R\left(\ul P_{i},\psi\left(x\right)\right),
\]
contradicting \eqref{eq:ineq_sumrank}. Hence, $\ul P$ must be minimal.
\end{itemize}
\end{proof}

\subsection{\label{subsec:pf_lem_diag}Proof of Lemma \ref{lem:Diagonal}}
\begin{proof}
Suppose not. Write
\begin{align*}
L & :=\left\{ i\in N:\ R\left(P_{i},x_{i}\right)\leq k\right\} ,\\
H & :=\left\{ i\in N:R\left(P_{i},x_{i}\right)\geq k+1\right\} .
\end{align*}
Then there is some $k\in\left\{ 1,...,N\right\} $ such that $\#\left(L\right)<k.$
As $\#\left(L\right)+\#\left(H\right)\equiv N$, then $\#\left(H\right)\geq N-k+1$.
Take any individual $h_{1}\in H$. and write
\[
Q_{h_{1}}:=\left\{ z\in M:z\succ_{P_{h_{1}}}x_{h_{1}}\right\} ,
\]
to denote the set of widgets that individual $h_{1}$ ranks higher
than $x_{h_{1}}$. By Lemma \ref{lem:NotAvail}, all widgets in $Q_{h_{1}}$
must have been assigned to someone in 
\[
N\backslash\left\{ h_{1}\right\} =L\cup\left(H\backslash\left\{ h_{1}\right\} \right).
\]
By construction, $\#\left(L\right)<k$ but
\begin{align*}
\#\left(Q_{h_{1}}\right) & =R\left(P_{h_{1}},x_{h_{1}}\right)-1\geq k,
\end{align*}
so there is at least one individual $h_{2}\in H\backslash\left\{ h_{1}\right\} $
with $x_{h_{2}}\in Q_{h_{1}},$i.e.,
\[
x_{h_{2}}\succ_{P_{h_{1}}}x_{h_{1}}.
\]
Now, by Lemma \ref{lem:NoCycle}, $h_{2}$ must also like her own
widget $x_{h_{2}}$ better than $h_{1}$'s widget $x_{h_{1}}$,
\begin{equation}
x_{h_{2}}\succ_{P_{h_{2}}}x_{h_{1}}.,\label{eq:ineq_h1_h2}
\end{equation}
otherwise it would be a Pareto improvement for $h_{1}$ and $h_{2}$
to exchange their widgets.

Consider
\[
Q_{h_{2}}:=\left\{ z\in M:z\succ_{P_{h_{2}}}x_{h_{2}}\right\} .
\]
with $\#\left(Q_{h_{2}}\right)\geq k$ again. Again, all widgets in
$Q_{h_{2}}$ must have been assigned to someone else. However, by
\eqref{eq:ineq_h1_h2}, $x_{h_{1}}\notin Q_{h_{2}}.$ Hence, all widgets
in $Q_{h_{2}}$ must have been assigned to some individual in
\[
N\backslash\left\{ h_{1},h_{2}\right\} =L\cup\left(H\backslash\left\{ h_{1},h_{2}\right\} \right).
\]
Again, as $\#\left(L\right)<k$, there exists at least one $h_{3}\in H\backslash\left\{ h_{1},h_{2}\right\} $
such that $h_{2}$ likes $h_{3}$'s widget better:
\[
x_{h_{3}}\succ_{P_{h_{2}}}x_{h_{2}},
\]
Now, by Lemma \ref{lem:NoCycle}, $h_{3}$ must like $x_{h_{3}}$
better than $x_{h_{2}}$ as well,
\[
x_{h_{3}}\succ_{P_{h_{3}}}x_{h_{2}}
\]
Moreover, by Lemma \ref{lem:NoCycle}, $h_{3}$ must also like $x_{h_{3}}$
better than $x_{h_{1}}$,
\[
x_{h_{3}}\succ_{P_{h_{3}}}x_{h_{1}},
\]
otherwise we could achieve a Pareto improvement by giving $x_{h_{2}}$
to $h_{1}$, $x_{h_{3}}$ to $h_{2}$ and $x_{h_{1}}$ to $h_{3}$.

We may carry out the same arguments inductively until the last element
$h_{\#\left(H\right)}$ in $H$ is reached within finite steps, as
$H$ is finite. By then we have enumerated all the elements in $H$
such that
\begin{align*}
x_{h_{l+1}}\succ_{P_{h_{l}}}x_{h_{l}} & ,\quad\forall l=1,...,\#\left(H\right)-1,
\end{align*}
and moreover, by Lemma \ref{lem:NoCycle},
\begin{equation}
x_{h_{\#\left(H\right)}}\succ_{h_{\#\left(H\right)}}x_{h_{l}},\quad\forall l=1,...,\#\left(H\right)-1.\label{eq:h_last}
\end{equation}
Define 
\[
Q_{h_{\#\left(H\right)}}:=\left\{ z\in M:z\succ_{P_{h_{\#\left(H\right)}}}x_{h_{\#\left(H\right)}}\right\} .
\]
By \eqref{eq:h_last}, for any $h^{'}\in H$, $x_{h^{'}}\notin Q_{h_{\#\left(H\right)}}.$
By Lemma \ref{lem:NotAvail}, all the widgets in $Q_{h_{\#\left(H\right)}}$
needs to be assigned to someone in $N\backslash H=L.$ As $\#\left(L\right)<k$
but $Q_{h_{\#\left(H\right)}}\geq k$, this is impossible. Hence,
we have reached a contradiction.
\end{proof}

\subsection{\label{subsec:pf_thm_private}Proof of Theorem \ref{thm:private}}
\begin{proof}
(a). Under the supposition, there exists an allocation $x\in X$ such
that
\[
R\left(\ol P_{i},x_{i}\right)=1,\quad\forall i\in N,
\]
which is the unique Pareto efficient allocation and achieves the best
possible ranking profile.

(b). Under $\ul P$, any Pareto efficient allocation must consist
of the unanimously agreed top $N$ widgets by Lemma \ref{lem:NotAvail}.
Hence, the set of Pareto efficient allocations consists exactly of
all permutations of the unanimously agreed top $N$ widgets among
the $N$ individuals. Moreover, for any Pareto efficient allocation
$x$, the ranking evaluation vector $R\left(\ul P,x\right)$ must
be a permutation of the vector:
\[
\ol r:=\left(1,2,....,n-1,N\right)^{'}.
\]

Now, consider any other preference profile $P$ and any Pareto efficient
allocation $x$ under $P$: $x\in PE\left(P\right)$. We now seek
to prove that there must exist a permutation mapping $\pi:\left\{ 1,...,N\right\} \to\left\{ 1,...,N\right\} $
such that $R\left(P,x\right)\leq\pi\left(\ol r\right).$ Without loss
of generality, we may permute individual indexes so that
\[
R\left(P_{1},x_{1}\right)\leq R\left(P_{2},x_{2}\right)\leq...\leq R\left(P_{N},x_{N}\right),
\]
and write 
\[
r:=\left(R\left(P_{1},x_{1}\right),\ R\left(P_{2},x_{2}\right),...,\ R\left(P_{N},x_{N}\right)\right)^{'}.
\]
Applying Lemma \ref{lem:Diagonal} with $k=1$, we have $r_{1}=1.$
Now applying Lemma \ref{lem:Diagonal} with $k=2$, we have
\begin{align*}
\#\left\{ i\in N:\ R\left(P_{i},x_{i}\right)\leq2\right\}  & \geq2,
\end{align*}
so we have $r_{2}\in\left\{ 1,2\right\} \leq2.$ Inductively, given
that $r_{i}\leq i$ for all $i=1,...,k-1$,we must have by $r_{k}\leq k$
by Lemma \ref{lem:Diagonal}(a). Hence, we have $r\leq\ol r.$ In
summary, there exists a permutation mapping $\pi_{x}:\left\{ 1,...,N\right\} \to\left\{ 1,...,N\right\} $
such that
\[
R\left(P,x\right)\leq\pi_{x}\left(\ol r\right).
\]

Now we construct the mapping $\psi$ from $PE\left(\ul P\right)$
to $PE\left(P\right)$.

For each $x\in PE\left(\ul P\right)$, there is a unique permutation
$\left(i_{1},....,i_{n}\right)$ of $N$ such that $R\left(P_{i_{k}},\ul x_{i_{k}}\right)=k.$
We now construct a Pareto efficient allocation $y$ under $P$ in
the following way. For $i_{1}$, we assign in $y$ to $i_{1}$ her
favorite widget in $M$ under $P_{i_{1}}$, i.e., $y_{i_{1}}:=R^{-1}\left(P_{i_{1}},1\right).$
Trivially,
\[
R\left(P_{i_{1}},y_{i_{1}}\right)=1\leq1=R\left(\ul P_{i_{1}},x_{i_{1}}\right).
\]
Inductively, given $y_{i_{1}},....,y_{i_{k-1}}$such that
\[
R\left(P_{i_{h}},y_{i_{h}}\right)\leq h=R\left(\ul P_{i_{h}},x_{i_{h}}\right),\quad\forall h=1,..,k-1,
\]
we assign in $x$ to $i_{k}$ her favorite widget in $M\backslash\left\{ y_{1},...,y_{k-1}\right\} $,
i.e.,
\[
y_{i_{k}}:=\arg\min_{m\in M\backslash\left\{ y_{i_{1}},...,y_{i_{k-1}}\right\} }R\left(P_{i_{k}},m\right).
\]
As $\left\{ y_{i_{1}},...,y_{i_{k-1}}\right\} $ involve only $k-1$
widgets, there must exist a widget $m\in M\backslash\left\{ y_{i_{1}},...,y_{i_{k-1}}\right\} $
such that $R\left(P_{i_{k}},m\right)\leq k$, implying that
\begin{equation}
R\left(P_{i_{k}},y_{i_{k}}\right)\leq k=R\left(\ul P_{i_{k}},x_{i_{k}}\right).\label{eq:ConstLower}
\end{equation}
Induction up to $k=N$ leads to a well-defined allocation $y$, and
we set $y=\psi\left(x\right)$.

First, note that $R\left(P,\psi\left(x\right)\right)\leq R\left(\ul P,x\right)$
by \eqref{eq:ConstLower}.

Second, $y$ is Pareto efficient under $P$. To see this, note that
each $i_{k}$ is getting her most preferred widget under $P$ among
all widgets \emph{not yet take}n by $i_{1},...,i_{k-1}$, implying
that no other allocations could strictly make $i_{k}$ better off
under $P$ without making one of the individuals among $i_{1},...,i_{k-1}$
worse off under $P$. Hence,
\[
\psi\left(PE\left(\ul P\right)\right)\subseteq PE\left(P\right).
\]

Finally, we show that $PE\left(P\right)\subseteq\psi\left(PE\left(\ul P\right)\right).$
Take any $y\in PE\left(P\right)$.

Clearly, there exists some $j_{1}\in N$ such that $R\left(P_{j_{1}},y_{j_{1}}\right)=1$
by Lemma \ref{lem:Diagonal}.

We now claim that there must exist some $j_{2}\in N\backslash\left\{ j_{1}\right\} $
such that
\[
y_{j_{2}}=\arg\min_{m\in M\backslash\left\{ y_{j_{1}}\right\} }R\left(P_{j_{2}},m\right).
\]
Suppose not. Taking any $k_{1}\in N\backslash\left\{ j_{1}\right\} $,
there must exist some $k_{2}\in N\backslash\left\{ j_{1}\right\} $
such that
\[
y_{k_{2}}=\arg\min_{m\in M\backslash\left\{ y_{j_{1}}\right\} }R\left(P_{k_{1}},m\right),
\]
which in particular implies that
\[
R\left(P_{k_{1}},y_{k_{2}}\right)<R\left(P_{k_{1}},y_{k_{1}}\right).
\]
Inductively we can find a sequence of individuals $k_{1},k_{2},...,k_{n}$
in $N\backslash\left\{ j_{1}\right\} $ such that
\[
R\left(P_{k_{h}},y_{k_{h+1}}\right)<R\left(P_{k_{h}},y_{k_{h}}\right)\text{ \ensuremath{\forall h\in\left\{ 1,...,N-1\right\} .}}
\]
However, as $\#\left(N\backslash\left\{ j_{1}\right\} \right)=N-1$,
so there must exist two $h_{1},h_{2}\in\left\{ 1,...,N\right\} $
such that $k_{h_{1}}=k_{h_{2}},$which contradicts with Lemma\ref{lem:NoCycle}.

Inductively, suppose we have constructed the sequence of individuals
$\left(j_{1},...,j_{k}\right)$ such that
\[
y_{j_{l}}=\arg\min_{m\in M\backslash\left\{ y_{j_{1}},...,y_{j_{l-1}}\right\} }R\left(P_{j_{l}},m\right)\text{ }\forall l=1,...,k.
\]
We claim that there must exist some $j_{k+1}\in N\backslash\left\{ j_{1},...,j_{k}\right\} $
such that
\[
y_{j_{k+1}}=\arg\min_{m\in M\backslash\left\{ y_{j_{1}},...,y_{j_{k}}\right\} }R\left(P_{j_{k+1}},m\right).
\]
Suppose not. Taking any $h_{1}\in N\backslash\left\{ j_{1},...,j_{k}\right\} $,
there must exist some $h_{2}\in N\backslash\left\{ j_{1},...,j_{k}\right\} $
such that
\[
y_{h_{2}}=\arg\min_{m\in M\backslash\left\{ y_{j_{1}},...,y_{j_{k}}\right\} }R\left(P_{h_{1}},m\right),
\]
which in particular implies that
\[
R\left(P_{h_{1}},y_{h_{2}}\right)<R\left(P_{h_{1}},y_{h_{1}}\right).
\]
Inductively we can find a sequence of not necessarily distinct individuals
$h_{1},h_{2},...,h_{n-k+1}$ in $N\backslash\left\{ j_{1},...,j_{k}\right\} $
such that
\[
R\left(P_{h_{l}},y_{h_{l+1}}\right)<R\left(P_{h_{l}},y_{h_{l}}\right)\text{ \ensuremath{\forall h\in\left\{ 1,...,N-k\right\} .}}
\]
However, as $\#\left(N\backslash\left\{ j_{1},...,j_{k}\right\} \right)=N-k$
while , so there must exist two $l_{1},l_{2}\in\left\{ 1,...,N-k+1\right\} $
such that $h_{l_{1}}=h_{l_{2}},$ which contradicts with Lemma\ref{lem:NoCycle}.

In summary, for each $y\in PE\left(P\right)$, we have constructed
a permutation $\left(j_{1},....,j_{n}\right)$ of $N$ such that
\[
y_{j_{k}}=\arg\min_{m\in M\backslash\left\{ y_{j_{1}},...,y_{j_{k-1}}\right\} }R\left(P_{j_{k}},m\right)\text{ }\forall k=1,...,N.
\]
For this given permutation $\left(j_{1},....,j_{N}\right)$, construct
an allocation $x$ by setting $x_{j_{k}}:=R^{-1}\left(\ul P,k\right),$
i.e., giving individual $j_{k}$ the $k$-th best widget under the
common preference $\ul P$. Then we have $y=\psi\left(x\right).$

Hence, we have constructed an onto mapping $\psi:PE\left(\ul P\right)\to PE\left(P\right)$
such that
\[
R\left(P,\psi\left(x\right)\right)\leq R\left(\ul P,x\right),\quad\forall x\in PE\left(\ul P\right).
\]
\end{proof}

\subsection{\label{subsec:pf_thm_IndMax}Proof of Theorem \ref{thm:IndMax}}
\begin{proof}
We present the proofs for (b)(c) below, as (a) has been already proved
in the main text.
\begin{itemize}
\item[(b)] Suppose that there exist some $i\in N$ and two distinct Pareto allocations
$x,y\in PE\left(P^{wel}\right)$ such that $x_{i}=y_{i}$ and $y_{-i}\succ_{P_{i,-i}^{wel}}x_{-i}$.
Then:
\[
y\in\left\{ z\in X:z_{i}=x_{i}\text{ and }z_{-i}\succ_{P_{i,-i}^{wel}}x_{-i}\right\} \text{\ensuremath{\neq\es}},
\]
and \eqref{eq:EqIndP} is violated, resulting in ``$\vartriangleright_{X}$''.
\item[(c)] Suppose that $n\geq3$, $\#\left(PE\left(P^{wel}\right)\right)\geq2$
and $\left(P_{i,-i}\right)_{i\in N}$ is strict. We claim that there
exists some individual $i$ and two distinct $x,y\in PE\left(P^{wel}\right)$
such that $x_{i}=y_{i}$. If this claim is true, as $\left(P_{i,-i}\right)_{i\in N}$
is strict, then either $x_{-i}\succ_{P_{i,-i}}y_{-i}$ or $y_{-i}\succ_{P_{i,-i}}x_{-i}$
is true, satisfying the condition in (a.i) and our conclusion follows.
We now prove the above claim:

By the proof of Theorem \eqref{thm:private}(b), for each individual
$i$, there exists at least one Pareto efficient allocation $x\in PE\left(P^{wel}\right)$
such that $R_{M}\left(P_{i,i},x_{i}\right)=1$. Now, suppose that
for each individual $i$, there exists just a single Pareto efficient
allocation $x^{\left(i\right)}\in PE\left(P^{wel}\right)$ such that
$R_{M}\left(P_{i,i},x_{i}^{\left(i\right)}\right)=1$. For each $i$
and $x^{\left(i\right)}$, $x_{-i}^{\left(i\right)}$ must be the
unique Pareto efficient allocation in $X_{-i}\backslash\left\{ z\in X:z_{i}=x_{i}^{\left(i\right)}\right\} $
according to $\left(P_{j,j}\right)_{j\neq i}$. This is only possible
if $x_{j}^{\left(i\right)}$ is $j$'s favorite widget in $M\backslash\left\{ x_{i}^{\left(i\right)}\right\} $
for every $j\neq i$. Hence, $R_{M}\left(P_{j,j},x_{j}^{\left(i\right)}\right)\leq2$
for every $j\neq i$. If $R_{M}\left(P_{j,j},x_{j}^{\left(i\right)}\right)=1$
for all $j\neq i$, then we must have $PE\left(P^{wel}\right)=\left\{ x^{\left(i\right)}\right\} $,
contradicting $\#\left(PE\left(P^{wel}\right)\right)\geq2$. Hence,
there exists at least one individual $j\neq i$ such that $R_{M}\left(P_{j,j},x_{j}^{\left(i\right)}\right)=2$.
Now, by Lemma \eqref{lem:NotAvail}, $j$'s favorite widget, denoted
$x_{j}^{*}$, must not be available, and in particular we must have
$x_{i}^{\left(i\right)}=x_{j}^{*}$. Now, as $n\geq3$, there must
exists another individual $k\neq i,j$.

If $x_{k}^{\left(i\right)}$ is individual $k$'s favorite widget,
namely $x_{k}^{*}$ in $M$, then there must exist another Pareto
efficient allocation $\ol x\in PE\left(P^{wel}\right)$ such that
$\ol x_{j}=x_{j}^{*}$ and $\ol x_{k}=x_{k}^{*}$ with $R_{M}\left(P_{j,j},\ol x_{j}\right)=R_{M}\left(P_{j,j},\ol x_{k}\right)=1$,
i.e., we give $i,j$'s common favorite widget $x_{j}^{*}$ to individual
$j$ and give individual $k$'s favorite widget to $k$. Then, we
have two distinct $x^{\left(i\right)},\ol x\in PE\left(P^{wel}\right)$
such that $x_{k}^{\left(i\right)}=\ol x_{k}=x_{k}^{*}$ for individual
$k$, proving our claim above.

Otherwise if $x_{k}^{\left(i\right)}$ is not individual $k$'s favorite
widget in $M$, then individual $k$'s favorite widget, denoted $x_{k}^{*}$,
must also be taken by individual $i$, i.e., $x_{i}^{\left(i\right)}=x_{k}^{*}$,
and in the meanwhile $R_{M}\left(P_{k,k},x_{k}^{\left(i\right)}\right)=2$.
Now, there must exist another Pareto efficient allocation $\ol x\in PE\left(P^{wel}\right)$
such that $\ol x_{j}=x_{j}^{*}$ and $\ol x_{k}=x_{k}^{\left(i\right)}$
with $R_{M}\left(P_{j,j},\ol x_{j}\right)=1$ and $R_{M}\left(P_{k.k},\ol x_{k}\right)=2$,
where we give $i,j,k$'s favorite widget $x_{j}^{*}$ to $j$ and
give $k$'s second favorite widget $x_{k}^{\left(i\right)}$ to $k$.
Again, we have two distinct allocations $x^{\left(i\right)},\ol x\in PE\left(P^{wel}\right)$
such that $x_{k}^{\left(i\right)}=\ol x_{k}=x_{k}^{*}$ for individual
$k$, proving our claim above.
\end{itemize}
\end{proof}

\end{document}